\documentclass[11pt]{article}
\usepackage{amssymb}
\usepackage{amsmath}
\usepackage{epsfig}
\usepackage{mathrsfs}
\usepackage[english]{babel}

\usepackage{xcolor}
\definecolor{darkblue}{rgb}{0.1,0.1,.7} 
\usepackage[breaklinks,colorlinks, linkcolor=darkblue]{hyperref}

\newcommand{\be}{\begin{equation}}
\newcommand{\ee}{\end{equation}}
\newcommand{\bea}{\begin{align}}
\newcommand{\eea}{\end{align}}
\newcommand{\nn}{\nonumber}

\newcommand{\A}{{\cal A}}
\newcommand{\B}{{\cal B}}

\newcommand{\D}{{\cal D}}
\newcommand{\M}{{\cal M}}
\newcommand{\R}{{\cal R}}
\newcommand{\OO}{{\cal O}}
\newcommand{\Y}{{\cal Y}}

\renewcommand{\d}{\delta}

\newcommand{\rmd}{{\rm d}}

\newcommand{\bt}[1]{{\bar t}}
\newcommand{\ts}{\textstyle}

\newcommand{\half}{{\ts \frac{1}{2}}}
\newcommand{\pr}{\partial}
\newcommand{\vep}{\varepsilon}
\newcommand{\vphi}{\varphi}
\newcommand{\tphi}{\tilde \varphi}
\newcommand{\tv}{\tilde v}
\newcommand{\tz}{\tilde z}
\newcommand{\ck}{c} 
\newcommand{\lt}{\left}
\newcommand{\rt}{\right}

\newcommand{\Oe}{${\rm O}(\vep)$}
\newcommand{\Oee}{${\rm O}(\vep^2)$}
\DeclareMathOperator{\tr}{tr}

\csname @addtoreset\endcsname{equation}{section}
\begin{document}

\begin{titlepage}
\thispagestyle{empty}
\begin{flushright}
\small
DAMTP-2007-77\\
arXiv:0708.2697\\
\date \\
\normalsize
\end{flushright}

\vskip 3cm
\centerline{\LARGE \bf{Epsilon Expansion for Multicritical Fixed Points}}
\vskip 6pt
\centerline{\LARGE \bf{and Exact Renormalisation Group Equations}}

\vskip 3cm

\centerline{J. O'Dwyer\footnote{jpo23@damtp.cam.ac.uk}$~$  and   
H. Osborn\footnote{ho@damtp.cam.ac.uk}} 
\vskip 1cm
\centerline{Department of Applied Mathematics and Theoretical Physics,} 
\centerline{Wilberforce Road, Cambridge, CB3 0WA, England}


\begin{abstract}
The Polchinski version of the exact renormalisation group equations
is applied to multicritical fixed points, which are present for dimensions
between two and four, for scalar theories using both the local potential
approximation and its extension, the derivative expansion. The results
are compared with the epsilon expansion by showing that the non linear differential
equations may be linearised at each multicritical point and the epsilon
expansion treated as a perturbative expansion. The results for critical
exponents are compared with corresponding epsilon expansion results from 
standard perturbation theory. The results provide a test for the validity
of the local potential approximation and also the derivative expansion.
An alternative truncation of the exact RG equation leads to equations
which are similar to those found in the derivative expansion but which
gives correct results for critical exponents to order $\varepsilon$ and also 
for the field anomalous dimension to order $\varepsilon^2$. An exact marginal
operator for the full RG equations is also constructed.

\end{abstract}

\vfill ${~~~}$ \newline

\end{titlepage}

\setcounter{footnote}0
\section{Introduction}

A fundamental development in quantum field theory was understanding 
the role of the renormalisation scale induced by the presence of a cut off, or
any other regularisation ensuring finiteness, and the associated flow of the
couplings of the theory under changes of scale. The RG flow equations
therefore reflect the essential arbitrariness of the renormalisation scale.
Nevertheless the global nature of the renormalisation flows in the space of 
couplings and the various fixed points that are present are crucial properties 
of any particular quantum field theory of physical interest, although in 
general their analysis is beyond the scope of conventional perturbation theory.
Since the time of Wilson \cite{Wilson,Wegner,Wilson3,Polchinski}
various exact RG equations have been formulated which in principle transcend
perturbation theory and allow the determination of fixed points and also the 
critical exponents that determine the flow of the couplings in the 
neighbourhood of fixed points, for recent reviews see 
\cite{Bagnuls,Wett,Jan,Bertrand} and for a critical discussion \cite{Dela}.

For theories involving just scalar fields, when a cut off function is introduced
in the quadratic part of the action,  these have been extensively explored.
At a rigorous level they may be used to provide an alternative proof of
the renormalisability of such theories \cite{Polchinski,Muller}. On the other
hand outside the perturbative domain it is necessary to resort to approximations
when the functional differential equations for the RG flow of the effective action,
which are in principle exact,
are reduced to non linear coupled differential equations 
which may then be analysed numerically. The simplest approximation is 
when the effective action, in general a nonlocal functional of local fields, is 
restricted to a function just of the scalar field without any derivatives, the 
local potential approximation (LPA) \cite{Hasenfratz}.  Beyond the LPA
it is possible to consider a derivative expansion to second and potentially
higher orders in the number of derivatives. However these approximations are
essentially uncontrolled. The resulting equations depend in detail on the form
of the cut off function and it is unclear whether there is any systematic 
procedure for improving, in principle, order by order the accuracy of results for 
critical exponents which should be independent of the particular form of the cut off.

Despite such difficulties the numerical results are often impressive and are
in good agreement with other methods of determining critical exponents for appropriate
statistical field theories in three dimensions. The LPA is applicable to various
different versions of the exact renormalisation group. In general the resulting
equations are inequivalent but 
the LPA for the Polchinski equation \cite{Polchinski} with  scalar fields, where 
the cut off dependence can be removed by simple rescalings and so is absent 
from calculated critical exponents, the results are  identical to 
the LPA ERG equations for the one particle irreducible generating function with a 
particular smooth cut off function \cite{Litim, Morris}. 
Expanding the action as an integral over local functions of the fields
with increasing numbers of derivatives then at other than zeroth order
there is an intrinsic dependence on the cut off in the resulting truncated equations
which cannot be removed by
redefinitions. For the Polchinski equation  this involves at each order just a finite 
set of parameters which are essentially arbitrary.

Nevertheless the basic LPA, yielding a simple nonlinear
differential flow equation for a potential $V(\phi)$, encapsulates the essential 
fixed point structure of such scalar theories. As the dimension $d$ is reduced 
a new fixed point is generated whenever the operator $\phi^{2n}$, for $n=2,3,\dots$,
becomes marginal. In the neighbourhood of each
fixed point the flow equations determine various critical exponents which may be
compared with results from other calculational methods.
A not yet fully realised goal is whether it is possible to improve the LPA,
while restricting to just a tractable finite set of coupled partial differential 
equations but with a systematic prescription for the determination of any
parameters present, so as to ensure that results for 
critical exponents should be quantitatively improved, closer to the results
of the particular quantum field theory, for all fixed points.

As a possible procedure for understanding how far the LPA and its extensions are 
valid we
consider here the connection with the $\vep$-expansion. As originally shown by
Wilson and Fisher \cite{Fisher} this provides a method whereby conventional quantum
field theory calculations of $\beta$-functions and related anomalous dimensions
as a loop expansion in $d=4-\vep$ dimensions may be applied to determine critical
exponents for $d=3$ as an asymptotic power expansion in $\vep$. For an extensive 
discussion in the context of standard quantum field theory see \cite{Hagen}. 
The $\vep$-expansion can also be obtained directly from exact RG equations, as was 
the case historically, since for $\vep\to 0$ the equations become linear and the 
non linear terms may be treated perturbatively. An interesting question
is then the extent to which the $\vep$-expansion results are compatible with those
from the LPA. Although this has been considered previously we here attempt a
systematic discussion in relation to the Polchinski RG equation. Initially
this is applied for just the LPA itself but we also consider derivative expansion
extensions to see whether any improvements in the domain of joint validity is
feasible. A similar discussion for $2<d<4$ is undertaken for the hierarchical
RG in \cite{Hier}.

An alternative approximation for the exact RG flow equations is to consider
expanding the effective action in terms of translation invariant functions of
the basic fields which are eigenfunctions of the linearised RG flow functional 
differential operator which are referred to as scaling fields \cite{Phase}. 
The non linear part of the RG flow equation may then be expanded in this basis.
This gives a set of coupled equations which in the simplest approximation is
equivalent to the LPA and at the next order is very similar to the derivative
approximation. However in this approach the dependence on the cut off function
is more controlled and in the $\vep$-expansion it is possible to get the
correct result for the critical exponent $\eta$ at order \Oee, unlike in the
usual derivative expansion. 

In this paper in Section \ref{sec:pertex} we first consider standard perturbative
calculations, with the aid of the background field method, for determining
critical exponents in the $\vep$-expansion at all multicritical points for
a single scalar field. This is applied both for scalar operators with no and
also two derivatives. Some higher order results, which involve multi-loop
calculations, are obtained in Appendix \ref{PVVV}. Although the methods used
are very different from exact RG calculations they provide results which are 
useful comparison for later approximations.  In Section \ref{LPAs} we consider
the LPA. It is shown how at \Oe\ the solution for each multicritical point is a 
single Hermite polynomial, whose coefficient is determined by the nonlinear terms,
and at \Oee\ it is just a finite sum. In Section \ref{App} the results are
worked out in more detail for the first three critical points and graphical
comparisons are made between the approximate analytic solution and numerical
solutions for various $d$. In \ref{sec:exponentsfirstorder} the corresponding
critical exponents, within the LPA, are found at \Oe\ and also \Oee, where
they disagree with the perturbative results.

The LPA is well known to be of restricted validity, it requires that the 
critical exponent $\eta$, which is essentially the anomalous dimension of the
elementary scalar field, is zero. The derivative expansion attempts to overcome
these limitations and we consider this in the context of the $\vep$-expansion
in Section \ref{sec:2od}. The coupled equations now depend on two cut off function
dependent constants $A,B$ but they now allow $\eta$ to be determined. 
The solutions in terms of Hermite polynomials may
also be extended to this case with some modifications. Following this in
\ref{sec:exponentssecondorder} we use these results to determine critical exponents
at \Oe\ for two classes of scalar operators. For one class the results are the same as
in the LPA case and agree with perturbation theory, for the other set of operators
which involve derivatives the calculated exponents depend on $A,B$.

The scaling field  approach based on the exact RG flow equation is considered in Section
\ref{Scaling}. A similar truncation to the derivative expansion is possible leading
to equations which also may be solved simply in the $\vep$-expansion. In this
case the dependence on the cut off function resides in various integrals. In special
cases these are independent of the precise cut off function and they then determine
universal results for critical exponents to \Oe\ and also $\eta$ to \Oee. The
relevant integrals are discussed in Appendix \ref{appB} where the cut off function
independent values are shown to be related to logarithmic divergences in two vertex
Feynman integrals. In Section \ref{mod} the resulting equations are recast as coupled
differential equations which are very similar, although different in detail, to those
arising in the derivative expansion. Some more general remarks are contained in a
conclusion. In Appendix \ref{appD} we obtain some exact results for perturbations of the
full RG flow equations and show how to construct an exact marginal operator. The existence
of such an operator, leading to a line of equivalent fixed points, ensures that the RG equations
determine $\eta$.

\section{Perturbation Calculations}
\label{sec:pertex}

We here discuss for the purposes of comparison a conventional quantum field
theory calculation of critical exponents at multicritical fixed points in
the $\vep$-expansion. We initially consider just the basic Lagrangian
\be
{\cal L}(\phi) = \half (\pr \phi )^2 + V(\phi) \, .
\label{lag}
\ee
For
\be
d = d_n - \vep \, , \qquad d_n = \frac{2n}{n-1} \, , \quad n=2,3,\dots \, 
\label{dee}
\ee
the theory is therefore 
renormalisable for $V(\phi)\equiv V(g,\phi)$ a polynomial of degree $2n$ and where $\{g\}$ are the couplings
parameterising $V$.  The counterterms necessary for finiteness, ${\cal L}_{\rm{c.t.}}(\phi)$, have just poles in $\vep$. 
For $\mu$ a regularisation scale
$\mu^{-\vep}\big ({\cal L}(\phi) + {\cal L}_{\rm{c.t.}}(\phi)\big ) = {\cal L}_0 (\phi_0) = 
\half (\pr \phi_0)^2 + V_0(\phi_0)$, $V_0(\phi) = V(g_0,\phi)$,  the usual perturbative  $\beta$-functions
and anomalous dimensions may be defined by
\be
\mu \frac{\rmd}{\rmd \mu} {\cal L}_0(\phi_0)  =  \bigg ( -  \vep 
- {\hat \gamma}_\phi \, \phi\frac{\pr}{\pr \phi} 
+ {\hat \beta}^V \! \cdot \frac{\pr}{\pr V} \bigg ) {\cal L}_0(\phi_0)  = 0 \, ,  
\label{RGdef}
\ee
with
\be
{\hat \beta}^V\! (\phi)  = V({\hat \beta}^g,\phi)\, , \qquad  {\hat \beta}^V \! \cdot \frac{\pr}{\pr V} = {\hat \beta}^g\cdot \frac{\pr}{\pr g} \, .
\ee
This implies
\be
\mu \frac{\rmd}{\rmd \mu} \, \phi \Big |_{g_0,\phi_0}  = - {\hat \gamma}_\phi\,  \phi\, , \quad
 {\hat \beta}^V\! (\phi)  =  \mu \frac{\rmd}{\rmd \mu} \, V(\phi) \Big |_{g_0,\phi}  + {\hat \gamma}_\phi \,  \phi V'(\phi) \, ,
\ee
where ${\hat \gamma}_\phi$ and ${\hat \beta}^V\! (\phi) $ can be decomposed as
\begin{align}
{\hat \gamma}_\phi
= {}& - \half \vep + \gamma_\phi \, , \nn \\{\hat \beta}^V \! (\phi) ={}&   \vep \big ( V(\phi)   - \tfrac12 \phi \,   V'(\phi)  \big ) +  \beta^V\! (\phi) =
 \vep \,V(\phi)  + {\hat \gamma}_\phi \,  \phi V'(\phi) +  {\tilde \beta}^V\! (\phi) \, ,
\label{betaV}
\end{align}
where  ${\tilde \beta}^V(\phi)$ depends just  on products of $V(\phi)$ with two or more derivatives.
For no mass scales other than $\mu$ there is a single dimensionless coupling $\lambda$ and
\be
V(\phi) \to  V_\lambda(\phi)= \frac{1}{(2n)!}\, \lambda \, \phi^{2n}  \, ,
\label{Vg}
\ee
so that
\be
{\hat \beta}^V\! (\phi)  \to  \frac{1}{(2n)!}\, {\hat \beta}^\lambda(\lambda) \,\phi^{2n}\, , \quad
 {\hat \beta}^\lambda(\lambda) =  - \vep\, (n-1) \lambda + \beta^\lambda(\lambda)\, , \quad  \gamma_\phi = \gamma_\phi(\lambda) \, .
 \label{Vred}
\ee
As  usual in the $\vep$-expansion, there may be fixed points where 
\be
{\hat \beta}^\lambda(\lambda_*)=0 \, , \qquad  \eta = 2\gamma_\phi(\lambda_*) \, ,
\label{fixP}
\ee
with $\lambda_*$ and the critical exponent $\eta$ expressible perturbatively as a power series in $\vep$. 

To determine the counterterms to ensure a finite theory it is sufficient as 
usual to consider connected one particle irreducible graphs. We adopt as the 
basic propagator
\be
G_0(x) = \frac{1}{4\pi} \, \frac{\Gamma(\nu)}{\pi^{\nu}} \, \frac{1}{(x^2)^\nu}\, ,
\qquad \nu = \half d - 1 \, ,
\ee
satisfying $-\pr^2 G_0(x) = \delta^d(x)$, and also use a background field approach, 
following similar methods used for four dimensional theories in \cite{JHo}, 
where
\be
\phi = \varphi + f \, ,
\ee
with $f$ the quantum field. Only vacuum graphs are then necessary and since 
with dimensional regularisation $G_0(x)|_{x=0} = 0$ no graphs with lines involving 
a single vertex need be included.

At lowest order for the one particle irreducible functional $W$ we have
\be
W_1 = \sum_{r \ge 2} \frac{1}{2 r!} \int \rmd^d x_1 \, \rmd^d x_2 \, 
V^{(r)}(\vphi_1) \, G_0(x_{12})^r \, V^{(r)}(\vphi_2) \, , \qquad
x_{ij} = x_i - x_j \, , \ \vphi_i = \vphi(x_i) \, .
\label{W1}
\ee
To evaluate this we note that
\be
\int \rmd^d x \, e^{i k\cdot x} \, G_0(x)^r = \frac{1}{(4\pi)^{r}} \,
\frac{\Gamma(\nu)^r}{\Gamma(r\nu)} \ \Gamma \big ( 1 - (r-1)\nu \big ) \,
\bigg ( \frac{k^2}{4\pi} \bigg )^{(r-1)\nu - 1} \, .
\label{div}
\ee
This has a pole whenever $(r-1)\nu = 1,2, \dots $ so that, assuming \eqref{dee},
it is easy to see from  \eqref{div} that for $\vep \to 0$
\begin{subequations}
\begin{align}
G_0(x)^n \sim {}& \frac{2}{\vep} \, \frac{1}{(4\pi)^n} \, 
\Gamma \Big( \frac{1}{n-1}\Big)^{n-1} \, \d^d(x) \, , \label{Gdiva} \\
G_0(x)^{2n-1} \sim {}&  \frac{1}{\vep} \, \frac{1}{(4\pi)^{2n}} \,
\frac{n-1}{n} \, \Gamma \Big( \frac{1}{n-1}\Big)^{2n-2} \, \pr^2 \d^d(x) \, .
\label{Gdivb}
\end{align}
\end{subequations}

{}From \eqref{W1}, assuming $r\le 2n$, the necessary counterterms are then
\begin{align}
{\cal L}_{\rm{c.t.}1} = {}& \frac{1}{\vep} \,  \frac{1}{(4\pi)^n} \, \frac{1}{n!}\,
\Gamma \Big( \frac{1}{n-1}\Big)^{n-1} \, V^{(n)}(\phi)^2  \nn \\
&{} - \frac{1}{\vep} \,  \frac{1}{(4\pi)^{2n}} \, \frac{n-1}{(2n)!}\, 
\Gamma \Big( \frac{1}{n-1}\Big)^{2n-2} \,  V^{(2n)}(\phi)^2 (\pr \phi)^2 \, ,
\label{Lct}
\end{align}
where the two terms arise from \eqref{Gdiva} and \eqref{Gdivb} at $n-1$ and $2n-2$ 
loops respectively. \eqref{Lct} then gives
\be
{\tilde \beta}_1^V\! (\phi) = \frac{1}{(4\pi)^n} \, \frac{n-1}{n!}\,
\Gamma \Big( \frac{1}{n-1}\Big)^{n-1} \, V^{(n)}(\phi)^2 \, , \quad
\beta^\lambda_1(\lambda) =  \frac{\lambda^2 }{(4\pi)^n} \, (n-1)\frac{(2n)!}{n!^3}\,
\Gamma \Big( \frac{1}{n-1}\Big)^{n-1} \, ,
\label{Vone}
\ee
and also, for the anomalous dimension 
$\gamma_\phi(\lambda)$ of the field $\phi$ which is non zero at $2(n-1)$ loops,
\be
\gamma_{\phi,1}(\lambda) = \frac{\lambda^2}{(4\pi)^{2n}} \, \frac{2(n-1)^2}{(2n)!}\,
\Gamma \Big( \frac{1}{n-1}\Big)^{2n-2} \, .
\label{gphi}
\ee

From \eqref{Vone} the fixed point \eqref{fixP} requires 
that
\be
\frac{\lambda_*}{(4\pi)^n} \, \frac{(2n)!}{n!^3}\,
\Gamma \Big( \frac{1}{n-1}\Big)^{n-1} = \vep \, ,
\label{gstar}
\ee
and hence  from \eqref{gphi}
\be
\eta = 4(n-1)^2 \, \frac{n!^6}{(2n)!^3} \, \vep^2  + {\rm O}(\vep^3) \, .
\label{etaone}
\ee

In order to analyse scalar operators formed by arbitrary powers of $\phi$  \eqref{Vred} is extended to
\be
V(\phi) =  V_\lambda(\phi)  + U(\phi)\, , \qquad
U(\phi)= \sum_k \frac{1}{k!} \, g_k \, \phi^k \, .
\label{Vextra}
\ee
Then
\begin{align}
{\hat \beta}^{V_\lambda+U}\! (\phi) = {}& \frac{1}{(2n)!}\, {\hat \beta}^\lambda(\lambda) \, \phi^{2n} +  \D_\lambda U(\phi) + {\rm O}(U^2) \, , \quad
\D_\lambda U(\phi) = \sum_k \frac{1}{k!} \, {\hat \gamma}_k(\lambda) \, g_k \, \phi^k \,, \nn \\
{\hat \gamma}_k(\lambda) = {}&  - \vep \, \half (k -2) + \gamma_k(\lambda) \, .
\label{bgp}
\end{align}
At a fixed point from \eqref{bgp} the scale dimensions for $\phi^k$ in the absence of mixing are given by
\be
\Delta_{\phi^k} = \frac{k}{n-1} - \vep + {\hat \gamma}_k(\lambda_*) \, .
\ee

The cases $k=1,2n-1$ are special.
In \eqref{betaV} ${\tilde \beta}^V$ for $V$ as in \eqref{Vextra} does not depend on $g_1$ so that
\be
{\hat \gamma}_1(\lambda)=  \vep+  {\hat \gamma}_\phi(\lambda) \, .
\label{gamma1}
\ee
We also have in general
\be
{\hat \gamma}_{2n-1} (\lambda)=  \frac{1}{\lambda} \,{\hat \beta}^\lambda(\lambda) -  {\hat \gamma}_\phi(\lambda) \, .
\label{gamman}
\ee
This follows since the contribution of $g_{2n-1}$ in \eqref{Vextra} is equivalent to differentiating the leading
${\rm O}(g)$ term and this extends to first order in $g_{2n-1}$ to differentiating the ${\hat \beta}^\lambda(\lambda) $ contribution to ${\hat \beta}^V\! (\phi) $
in \eqref{bgp} with respect to $\phi$  except where $\phi$ appears explicitly in \eqref{betaV} rather than in terms of $V(\phi)$ or its derivatives. 
At a fixed point we then have for the scaling dimensions 
\be
\Delta_\phi = \tfrac12 ( d-2+\eta)\, , \qquad \Delta_{\phi^{2n-1}} = \tfrac12 ( d+2-\eta)\, .
\ee

 Applying \eqref{betaV} and \eqref{Vone}  for \eqref{Vextra} gives to first order in $\lambda$
\be
\gamma_{k,1}(\lambda) = \frac{2\lambda}{(4\pi)^n} \, \frac{n-1}{n!^2}\,
\Gamma \Big( \frac{1}{n-1}\Big)^{n-1} \, \frac{k!}{(k-n)!} \, .
\label{gamk}
\ee
and then the result \eqref{gamk} then gives to first order in $\vep$
\be
{\hat \gamma}_{k,1}(\lambda_*) = - \half ( k - 2) \vep + 2(n-1)\, \frac{n!}{(2n)!} \, 
\frac{k!}{(k-n)!} \, \vep \, .
\label{expk}
\ee

The $\vep$-expansion at multicritical points using standard quantum field
theory was considered in \cite{Itzykson} who obtained \eqref{gstar} and
\eqref{etaone}. In three dimensions, corresponding to $n=3$, results equivalent
to \eqref{gamk} for $k=1,\dots,5$ were obtained in \cite{Toms}.

The local operators in the basic quantum field include also those with
derivatives as well as just $\phi^k$. These are relevant for scalar operators when $k\ge 2n$. 
To extend the above discussion we
consider in addition to \eqref{lag} ${\cal L}\to {\cal L} + {\cal L}^Z $ where
\be
{\cal L}^Z = Z(\phi) \pr^2 \phi \, , \qquad
Z(\phi) = \sum_{k\ge 2n} \frac{1}{(k-2n+1)!} \, h_k \, \phi^{k-2n+1} \, ,
\label{lagZ}
\ee
where we keep only graphs which involve one $Z$-vertex. 
With $\mu^{-\vep}({\cal L}^Z + {\cal L}^Z_{\rm{c.t.}} ) = {\cal L}_0^Z$
the corresponding $\beta$-function is given by
\be
\mu \frac{\rmd}{\rmd \mu} {\cal L}^Z \Big |_{{\cal L}_0+{\cal L}_0^Z} = 
- {\hat \gamma}_\phi \, \phi\frac{\pr}{\pr \phi} {\cal L}^Z +
 {\hat \beta}^Z (\phi) \pr^2 \phi \, , \quad 
{\hat \beta}^Z (\phi) =  \vep  \half \big ( Z(\phi)-\phi Z'(\phi)\big )  + {\beta}^Z (\phi) \, ,
\ee
which can be expanded as
\begin{align}
{\hat \beta}^Z (\phi) 
= {}& \sum_{k \ge 2n} \frac{1}{(k-2n+1)!} \, \big (
{\hat \gamma}^{hh}_k(\lambda) \, h_k + {\gamma}^{hg}_k(\lambda)\, g_k \big )\, \phi^{k-2n+1} 
\, , \nn \\
& {\hat \gamma}^{hh}_k(\lambda) = - \vep \, \half (k-2n) + \gamma^{hh}_k(\lambda)\, .
\label{betaZ}
\end{align}
We must also extend \eqref{bgp} to include mixing effects if $k\ge 2n$ to the
form
\be
{\hat \beta}^V\! (\phi) = \frac{1}{(2n)!}\, {\hat \beta}^\lambda(\lambda) \, \phi^{2n} +
\sum_k \frac{1}{k!} \, \big ( {\hat \gamma}^{gg}_k(\lambda) \, g_k 
+ {\gamma}^{g h}_k(\lambda) \, h_k \big )\, \phi^k \, , \quad {\hat \gamma}^{gg}_k(\lambda) 
= {\hat \gamma}_k(\lambda) \, .
\label{bgp2}
\ee
The terms in ${\cal L}^Z$ involving $g_{2n},h_{2n}$ may be absorbed in ${\cal L}$
by a redefinition of $\phi,\lambda$ giving for this special case
\begin{align}
{\hat \gamma}^{gg}_{2n}(\lambda) = {}& {\hat \beta}^\lambda{}'(\lambda) - 
2n\lambda\, {\hat \gamma}_\phi{}'(\lambda) \, ,
\quad \gamma^{gh}_{2n}(\lambda) = 2n\lambda \, {\hat \gamma}^{gg}_{2n}(\lambda) -
2n\, {\hat \beta}^\lambda(\lambda) \, ,\nn \\
\gamma^{hg}_{2n}(\lambda) = {}&  {\hat \gamma}_\phi{}'(\lambda) \, , \hskip 2.53cm
{\hat \gamma}^{hh}_{2n}(\lambda) = 2n\lambda\,  {\hat \gamma}_\phi{}'(\lambda) \, .
\label{spec}
\end{align}

The anomalous dimensions of operators at the fixed point are then given by the 
eigenvalues
$\omega_{k,1}, \, \omega_{k,2}$ of the matrix 
\begin{equation}
\Gamma_k=\lt(\begin{array}{cc} {\hat \gamma}^{gg}_k(\lambda_*) & {\gamma}^{gh}_k(\lambda_*)  \\
\noalign{\vskip 3pt}
{\gamma}^{hg}_k(\lambda_*) & {\hat \gamma}^{hh}_k(\lambda_*)  \\ \end{array}\rt) \, ,
\label{anom}
\end{equation}
for $k=2n,\dots,4n-3$.
It is easy to see from \eqref{spec} that 
\begin{equation}
\Gamma_{2n}=\lt(\begin{array}{cc} {\hat \gamma}^{gg}_{2n}(\lambda_*)   &  2n\lambda_* \, {\hat \gamma}^{gg}_{2n}(\lambda_*)  \\
\noalign{\vskip 3pt}
 {\hat \gamma}_\phi{}'(\lambda_*) \,& 2n\lambda_*\,  {\hat \gamma}_\phi{}'(\lambda_*)  \\ \end{array}\rt)
\quad \Rightarrow \quad \det \Gamma_{2n} =0  \, , \ \  \tr  \Gamma_{2n} =  {\hat \beta}^\lambda{}'(\lambda_*) \, .
\label{anom2}
\end{equation}
Hence the fixed point eigenvalues are
\be
\omega_{2n,1} = {\hat \beta}^\lambda{}'(\lambda_*)\, , \quad \omega_{2n,2} =0\, .
\ee
More generally, by using the equations of motion,
$\omega_{k,2}= \omega_{k-2n+1,1}-\half \vep -  \half \eta$ implying $\Delta_{k,2} = \Delta_{k-2n+1,1} + d - \Delta_\phi$.
 \eqref{gamma1} ensures
$\omega_{2n,2} =0$.

Since with dimensional regularisation
\be
G_0(x)^n \pr^2 G_0(x) = 0 \, , \quad n\ge 1\, ,
\ee
so that, to first  order in $V$,  the contributions involving $Z$ in addition to \eqref{W1} are just
\begin{align}
W^{Z}_1 =  \sum_{r\ge 2} \frac{1}{ r!}& \int \rmd^d x_1 \, \rmd^d x_2 \,  V_\lambda{\!} ^{(r)}(\vphi_1)
  \, G_0(x_{12})^r\,  Z^{(r)}(\vphi_2) \pr^2 \vphi_2  \, .
\label{WZ1}
\end{align}
Using \eqref{Gdiva} shows that in addition to \eqref{Lct} the required counterterms are
\begin{align}
{\cal L}^Z_{\rm{c.t.}1} = {}& \frac{2}{\vep} \,  \frac{1}{(4\pi)^n} \, \frac{1}{n!}\,
\Gamma \Big( \frac{1}{n-1}\Big)^{n-1} \, 
 V_\lambda{\!}^{(n)}(\phi)\, Z^{(n)}(\phi)\pr^2 \phi   \, .
\label{LZct}
\end{align}
Since $\beta^Z_1 = (n-1)\vep {\cal L}^Z_{\rm{c.t.}1}$ then from \eqref{LZct}
we have at lowest order
\be
{\gamma}^{hh}_{k,1}(\lambda) = \frac{2\lambda}{(4\pi)^n} \, \frac{n-1}{n!^2}\,
\Gamma \Big( \frac{1}{n-1}\Big)^{n-1} \frac{(k-2n+1)!}{(k-3n+1)!} = \gamma_{k-2n+1,1}(\lambda)\, ,
\ee
and ${\gamma}^{gg}_{k,1}(\lambda) = {\gamma}_{k,1}(\lambda)$ as in \eqref{gamk}. This gives
to ${\rm O}(\vep)$
\be
{\hat \gamma}^{hh}_{k,1}(\lambda_*) = -\tfrac12(k-2n) \vep + 2(n-1) \,\frac{n!}{(2n)!}\,
\frac{(k-2n+1)!}{(k-3n+1)!}\, \vep \, .
\label{expkh}
\ee

For the off diagonal parts of the anomalous dimension matrix 
we may note that using \eqref{Gdivb}  implies a $2(n-1)$-loop contribution to $\beta^Z$
\be
\Delta \beta^Z_1(\phi)  =  \frac{1}{(4\pi)^{2n}} \, \frac{4(n-1)^2}{(2n)!}\,
\Gamma \Big( \frac{1}{n-1}\Big)^{2n-2} \,  V_\lambda{\!}^{(2n)}(\phi)\,  U^{(2n-1)}(\phi)  \, .
\ee
At this order $Z(\phi)$ does not generate additional contributions to $V_{\rm c.t.}(\phi)$ so that
\begin{align}
{\gamma}^{gh}_{k,1}(\lambda) = 0 \, , \qquad
{\gamma}^{hg}_{k,1}(\lambda) =  \frac{\lambda}{(4\pi)^{2n}} \, \frac{4(n-1)^2}{(2n)!}\,
\Gamma \Big( \frac{1}{n-1}\Big)^{2n-2} = 2 \, \gamma_{\phi,1}(\lambda)/\lambda \, .
\label{ghg1}
\end{align}
Hence to first order in $\vep$
\begin{equation}
\Gamma_{k,1} =\lt(\begin{array}{cc} {\hat \gamma}^{gg}_{k,1}(\lambda_*) & 0 \\
\noalign{\vskip 3pt}
{\gamma}^{hg}_{k,1} (\lambda_*) & {\hat \gamma}^{hh}_{k,1}(\lambda_*)  \\ \end{array}\rt)   \, ,
\label{anom3}
\end{equation}
 so that the lowest order eigenvalues
of \eqref{anom} are given by \eqref{expk} for $\omega_{k,1}$ and
\begin{align}
\omega_{k,2} 
={}&  - \half (k-2n)\vep + 2(n-1)\, \frac{n!}{(2n)!} \, 
\frac{(k-2n+1)!}{(k-3n+1)!} \, \vep + {\rm O}(\vep^2) \nn \\
= {}& \omega_{k-2n+1,1} -  \half \vep + {\rm O}(\vep^2) \, .
\label{expk2}
\end{align}
 
Further perturbative results for $\beta^V$ and anomalous dimensions are obtained in 
Appendix \ref{PVVV}.

\section{Local Potential Approximation}\label{LPAs}

In the LPA, the Polchinski RG equation in
$d$ dimensions  may be reduced  by an appropriate rescaling to the following 
renormalisation flow for a potential $V(\phi,t)$, 
\begin{equation}
\dot{V}(\phi,t)=V''(\phi,t)-V'(\phi,t)^2+dV(\phi,t)-\half(d-2)\, \phi V'(\phi,t) \, ,
\label{eq:polch}
\end{equation}
where $t=-\log\Lambda$ and $\Lambda$ is a cut off scale. 
At a fixed point, $V(\phi,t)\to V_*(\phi)$ which solves
\be
V_*''(\phi)-V_*'(\phi)^2+dV_*(\phi)-\half(d-2)\, \phi V_*'(\phi) =0 \, .
\label{eq:lpanonlin}
\ee
This equation has been extensively analysed, both numerically and 
analytically \cite{Felder,Lima}. There are two trivial solutions $V_*=0$, 
the Gaussian fixed point, and $V_*(\phi)= \half \phi^2 - \frac{1}{d}$, the so
called high temperature fixed point.
For non trivial solutions even in $\phi$, with $V_*{\!}'(0)=0$, 
and bounded below, it is necessary to fine tune $V_*(0)$ to ensure 
that there are  no singularities for all $\phi$. Such solutions appear
whenever $d$ is reduced below $2n/(n-1)$ for $n=2,3,\dots$.

For our purposes it is convenient to consider a further rescaling by defining
\begin{equation}
x=\half(d-2)^{\frac{1}{2}}\phi \, , \qquad v(x,t)=V(\phi,t) \, , \ \ 
v_*(x) = V_*(\phi) \, .
\label{eq:redef}
\end{equation}
Then \eqref{eq:lpanonlin} becomes
\begin{equation}
v_*''-2x v_*'+\frac{4d}{d-2}\, v_*=v_*'^2 \, ,
\label{eq:lpanonlin1.5}
\end{equation}
where $v_*'=\frac{\rmd v_*}{\rmd x}$.

We first consider the linearised form of \eqref{eq:lpanonlin1.5},
\begin{equation}
\frac{\rmd^2 v}{\rmd x^2}-2x\frac{\rmd v}{\rmd x}+ \frac{4d}{d-2}\, v=0 \, ,
\label{eq:lpalin}
\end{equation}
which becomes a valid approximation when $v(x)$ is small. 
Global solutions for all $x$
which are bounded by a power for large $x$ are only possible for  Hermite
polynomials $H_n(x)$, $n=0,1,2,\dots$ which satisfy
\begin{equation}
(D + n )  H_n =0 \, , \qquad 
D = \half \frac{\rmd^2}{\rmd x^2} - x \frac{\rmd }{\rmd x}  \, .
\label{Dop}
\end{equation}
Hence \eqref{eq:lpalin} has solutions with the appropriate behaviour
\begin{equation}
v(x)=cH_n(x)
\end{equation}
only when $d$ is restricted to
\begin{equation}
n=\frac{2d}{d-2}.
\end{equation}
With the further requirement that $v(x)$ be bounded below, we must restrict
to $n$  to be even and $c>0$. Relabelling $n\rightarrow 2n$ we then have
solutions of the linearised equation
\begin{equation}
v(x) =cH_{2n}(x) \quad \hbox{for} \quad d=d_n \, ,
\label{dep}
\end{equation}
with $d_n$ defined in \eqref{dee}.
This may then used as a starting point for the analysis of the full non linear
fixed point equation.

For the non-linear equation \eqref{eq:lpanonlin1.5} we consider the case
when $d$ is close to the value in \eqref{dep}, where the linearised solution
holds, and may be written as in \eqref{dee}. 
We therefore seek solutions as $\vep \to 0$ of the form
\begin{equation}
v_*(x)=  v_n(x) + {\rm O}(\vep^2) \, , \qquad
v_n(x) = \ck_n\vep H_{2n}(x) \, .
\label{vnn}
\end{equation}
The right hand side of  \eqref{eq:lpanonlin1.5} is clearly \Oee. Writing
\be
\frac{4d}{d-2}=4n+2(n-1)^2\vep+(n-1)^3\vep^2+ {\rm O}(\vep^3).
\label{eq:expansion1}
\ee
the leading \Oe\ terms on the left hand side of  \eqref{eq:lpanonlin1.5} are
then absent.  To \Oee\ we may determine $\ck_n$ in \eqref{vnn} by noting that
the differential operator $D + 2n $, which is hermitian with respect to the
measure $\rmd x \, e^{-x^2}$,  generates only functions orthogonal to $H_{2n}$ so
we must require at this order
\be
2(n-1)^2\vep \int_{-\infty}^{\infty}\!\!\!\!\rmd x \, e^{-x^2}H_{2n}(x) v_n (x)
=  \int_{-\infty}^{\infty}\!\!\!\!\rmd x \, e^{-x^2} H_{2n}(x)v_n{\!}'(x)^2 \, .
\label{nonl}
\ee
Both sides  may be evaluated using the integrals for Hermite polynomials
\begin{equation}
N_k \equiv \int_{-\infty}^{\infty}\!\!\!\! \rmd x \, e^{-x^2}{H_{k}(x)}^2
= 2^k\pi^{\frac{1}{2}}\, k! \, ,
\label{eq:Nm}
\end{equation}
and also for three $H_n$'s,
\begin{align}
G_{klm} \equiv {}& \int_{-\infty}^{\infty}\!\!\!\! \rmd x \,e^{-x^2} 
{H_{k}(x)}{H_{l}(x)}{H_{m}}(x)
=2^{s}\pi^{\frac{1}{2}}\frac{k!\,l!\,m!}{(s-k)!(s-l)!(s-m)!} \, , \nn \\
&{} s = \half ( k+l+m) \, , \qquad k,l,m \le s \, , 
\label{eq:Gklm}
\end{align}
with $s$ required to be an integer.
On the right hand side of \eqref{nonl}
\begin{align}
\int_{-\infty}^{\infty}\!\!\!\! \rmd x \, e^{-x^2} H_{2n}(x)v_n{\!}'(x)^2 & =
\ck_n{\!}^2\vep^2\int_{-\infty}^{\infty}\!\!\!\! \rmd x \,
e^{-x^2}H_{2n}(x){H_{2n}{\!}'(x)}^2\nn\\
& =-\half \ck_n{\!}^2\vep^2\int_{-\infty}^{\infty}\rmd x \,
{H_{2n}(x)}^2\frac{\rmd}{\rmd x}\big (e^{-x^2}H_{2n}{\!\!}'(x) \big ) \nn\\
& = 2n\ck_n{\!}^2\vep^2\int_{-\infty}^{\infty}\!\!\!\! \rmd x \, 
e^{-x^2} {H_{2n}(x)}^3 = 2n\ck_n{\!}^2\vep^2G_{2n\,2n\,2n} \, ,
\label{eq:RHS}
\end{align}
using standard identities. Hence \eqref{nonl} determines a value for $\ck_n$
\be
2(n-1)^2\ck_nN_{2n} = 2n\ck_n{\!}^2\, G_{2n\,2n\,2n}\quad
\Rightarrow \quad \ck_n =\frac{(n-1)^2}{2^nn} \,\frac{n!^3}{(2n)!^2} \, . 
\label{eq:kn}
\ee
Since $\ck_n>0$ then for relevant solutions in \eqref{vnn} we must have $\vep>0$.

We may now extend the solution \eqref{vnn} to \Oee\ by assuming the form
\be
v_*(x) = \ck_n\vep H_{2n}(x) + \ck_n\vep^2\sum_{m}a_mH_m(x) + {\rm O}(\vep^3) \, .
\label{vnext}
\ee
Inserting in \eqref{eq:lpanonlin1.5} with \eqref{eq:expansion1} and keeping only
terms which are \Oee\ gives
\be
(n-1)^2  H_{2n}(x) + \sum_{m} (4n-2m) a_m H_m(x)
= \ck_n H_{2n}{\!}'(x)^2 \, ,
\ee
and hence using the orthogonality properties of Hermite polynomials
we may determine $a_m$ for $m \ne 2n$,
\be
N_m(4n-2m)a_m= \ck_n \int_{-\infty}^{\infty}\!\!\!\!\rmd x \, 
e^{-x^2} H_m(x) H_{2n}{\!}'(x)^2 \, .
\ee
The integral on the right hand side may be calculated by using $H_n{\!}'=2n
H_{n-1}$ or with the following judicious integrations by parts
\begin{align}
\int_{-\infty}^{\infty}\!\!\!\! \rmd x \, e^{-x^2} H_m {H_{2n}{\!}'}^2 &
=-\frac{1}{2} \int_{-\infty}^{\infty}\!\!\!\! \rmd x \, e^{-x^2}
H_m{\!}'\frac{\rmd}{\rmd x}\big (H_{2n}{\!}^2 \big )
-\int_{-\infty}^{\infty}\!\!\!\! \rmd x \, 
H_m H_{2n} \frac{\rmd}{\rmd x}\big (e^{-x^2}H_{2n}{\!}' \big )  \nn\\
& =\frac{1}{2} \int_{-\infty}^{\infty}\!\!\!\! \rmd x \,
\frac{\rmd}{\rmd x}\big (e^{-x^2} H_m{\!}' \big )H_{2n}{\!}^2
+4n\int_{-\infty}^{\infty}\rmd x \, e^{-x^2}H_mH_{2n}{\!}^2 \nn\\
& =(4n-m)\int_{-\infty}^{\infty}\!\!\!\! \rmd x \, e^{-x^2}H_mH_{2n}{\!}^2\nn\\
& =(4n-m)\, G_{m\,2n\,2n} \, ,
\end{align}
which  is zero unless $m$ is even. Hence taking $m=2p$,
\begin{align}
a_{2p}& =\frac{\ck_n}{N_{2p}}\,\frac{2n-p}{2(n-p)}\, G_{2p\,2n\,2n}\nn\\
& =  2^{n-p-1} \frac{(n-1)^2}{n}\, \frac{2n-p}{n-p} \, 
\frac{n!^3}{p!^2(2n-p)!} \, , \quad p=0,1,\dots 2n-1, \, p\ne n\, .
\label{eq:a2p}
\end{align}

Finally we compute $a_{2n}$ by using \eqref{vnext} in \eqref{eq:lpanonlin1.5}
and imposing orthogonality with $H_{2n}$ to  ${\rm O}(\vep^3)$, just as $\ck_n$ was 
determined in \eqref{nonl}. 
\begin{align}
(n-1)^2a_{2n}N_{2n}\ &+\ \half (n-1)^3 N_{2n}\nn\\
& = \ck_n\sum_{p} a_{2p}\int_{-\infty}^\infty \!\!\!\! \rmd x \, 
e^{-x^2}H_{2n}H_{2n}{\!}'H_{2p}{\!}'  \nn \\
& = 2 \ck_n\sum_{p}pa_{2p}\int_{-\infty}^\infty \!\!\!\! \rmd x \, 
e^{-x^2}H_{2n}{\!}^2 H_{2p}  \nn\\
& = 2\ck_n\sum_{p\neq n}pa_{2p}G_{2n\,2n\,2p}+
2 n\ck_na_{2n}G_{2n\,2n\,2n}
\label{eq:a2n}.
\end{align}
Using \eqref{eq:kn}, \eqref{eq:a2n} becomes
\begin{align}
(n-1)^2a_{2n}N_{2n}= \half (n-1)^3 N_{2n}- 2\ck_n\sum_{p\neq
n}pa_{2p}G_{2n\,2n\,2p},
\end{align}
and so
\begin{equation}
a_{2n}=\frac{1}{2}(n-1) -\frac{2\ck_n}{(n-1)^2}\sum_{p\neq n}
p\,a_{2p}\,2^p\frac{(2n)!(2p)!}{p!^2(2n-p)!}
\label{eq:a2n2},
\end{equation}
where we may use~\eqref{eq:kn} and~\eqref{eq:a2p}.

Since $H_n(x)\sim 2^n x^n$ for large $x$ then for $\vep x^2 = {\rm O}(1)$
the next to leading terms in \eqref{vnext} are comparable with the leading 
$H_{2n}(x)$ term so the $\vep$ expansion for $v_*(x)$ breaks down. 
From \eqref{eq:a2p} $a_{2p}<0$ for $p>n$ so that the result given
by \eqref{vnext}, which is a polynomial of degree $4n-2$, is negative 
for sufficiently large $x$. For the exact solution to \eqref{eq:lpanonlin1.5}
the nonlinear terms play a crucial role for large $x$ and we have
$v_*(x) \sim \frac{2}{d-2}x^2$.

\section{Applications in Particular Cases}\label{App}

The results obtained for $v_*$ obtained above are here considered in more
detail for $n=2,3,4$ and compared with results of numerical calculations.
In particular $V_*(\phi)$ was calculated numerically in the LPA in \cite{HarveyFros}
for these cases and various values of the dimension $d$.
The basic approximation from \eqref{vnext} is 
\begin{equation}
v_*(x)=\ck_n\vep
H_{2n}(x)+\ck_n\vep^2\sum_{p=0}^{2n-1} {a}_{2p}H_{2p}(x)
\label{eq:fp},
\end{equation}
where $a_{2p}$ is given by \eqref{eq:a2p} and \eqref{eq:a2n2}.

The first case of interest is $n=2$, corresponding to the Wilson-Fisher 
fixed point.
From~\eqref{eq:a2p} we have
\begin{equation}
k_2 = \frac{1}{4!^2} \, , \qquad
{a}_{2p}= \frac{4-p}{2-p} \, \frac{2^{3-p}}{p!^2(4-p)!} \, , \quad p= 0,1,3 \, ,
\end{equation}
and hence
\be
{a}_{0}=\frac{2}{3}\, , \quad
{a}_{2}=2\,, \quad
{a}_{6}= -\frac{1}{36} \, .
\ee
{}From these, and~\eqref{eq:a2n2}, we have that
\be
{a}_4 
= \frac{3}{2} \, .
\ee

Using the following results for Hermite polynomials,
\begin{align}
H_{2n}(0)=(-1)^n\frac{(2n)!}{n!} \, , \qquad H''_{n}(0)=-2nH_n(0) \, ,
\end{align}
we may then obtain from \eqref{eq:fp} in this case
\be
v_*(0)  =\frac{1}{48}\vep \lt( 1 +\frac{3}{2}\vep\rt) \, ,  \qquad
v''_*(0) = - \frac{1}{6}\vep\lt(1+\frac{7}{4}\vep\rt) \, .
\ee
As a consequence of \eqref{eq:redef} for the original $V_*(\phi)$ we have
$V_*(0)= v_*(0)$ and 
\begin{align}
V_*''(0)& = {\ts\frac{1}{4}}(d-2) v_*''(0) \nn \\
& =-\frac{\vep}{12}\lt(1+\frac{5\vep}{4}\rt) + {\rm O}(\vep^3) \, .
\label{Vpp2}
\end{align}
This results satisfy the consistency check
\begin{equation}
\frac{V_*''(0)}{V_*(0)} = -4\lt(1-\frac{1}{4} \vep\rt)=-d \, ,
\end{equation}
as follows directly from \eqref{eq:lpanonlin}. As remarked earlier
for solutions of the fixed point equation without singularities it
is necessary to fine tune $V_*''(0)$, or equivalently $V_*(0)$.
In Table~\ref{comparison1} we compare the results from \eqref{Vpp2} to \Oee\
with those from numerical calculation, as contained in \cite{HarveyFros},
for various $d$. 
The detailed form of the approximate solution in comparison with numerical
results is shown in Figure 1 for various $d$.
For small $\vep$ the agreement is good.
\begin{table}[!h]
\begin{tabular}{|r|r|r|} \hline
$d$&Numerical&$\vep$-expansion at \Oee\\
\hline
4&\hfil$0$~~&\hfil$0\qquad$\\
3.9&\hfil$-0.009$~~&\hfil$-0.009\qquad$\\
3.8&\hfil$-0.021$~~&\hfil$-0.021\qquad$\\
3.7&\hfil$-0.035$~~&\hfil$-0.034\qquad$\\
3.6&\hfil$-0.051$~~&\hfil$-0.050\qquad$\\
3.5&\hfil$-0.070$~~&\hfil$-0.068\qquad$\\
3.4&\hfil$-0.092$~~&\hfil$-0.088\qquad$\\
3.3&\hfil$-0.119$~~&\hfil$-0.109\qquad$\\
3.2&\hfil$-0.149$~~&\hfil$-0.133\qquad$\\
3.1&\hfil$-0.186$~~&\hfil$-0.159\qquad$\\
3.0&\hfil$-0.229$~~&\hfil$-0.188\qquad$\\
\hline
\end{tabular}\caption{Comparison of ERG numerical and analytical results for
$V_*''(0)$ for the case $n=2$.}\label{comparison1}
\end{table}
\begin{figure}[!h]
\vskip -6pt{
\hskip -1cm{
\scalebox{0.70}{\includegraphics{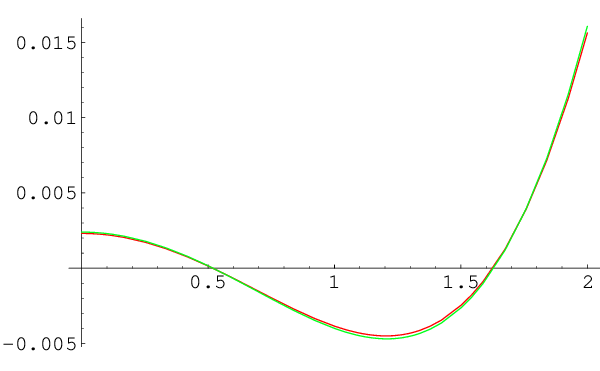}}
\scalebox{0.70}{\includegraphics{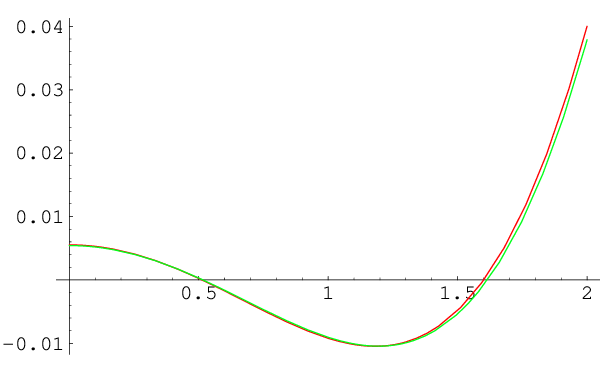}}}}
\vskip -6pt{
\hskip -1cm{
\scalebox{0.70}{\includegraphics{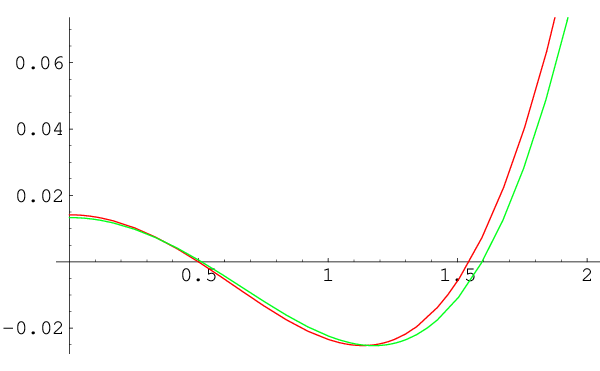}}
\scalebox{0.70}{\includegraphics{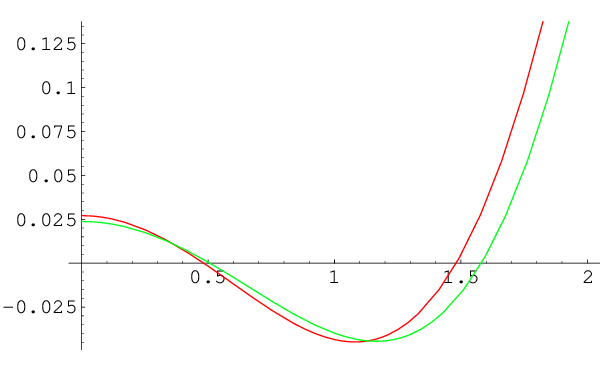}}}}
\vskip -6pt{
\hskip -1cm{
\scalebox{0.70}{\includegraphics{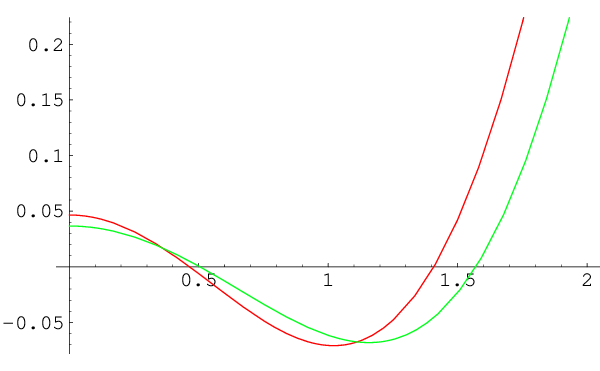}}
\scalebox{0.70}{\includegraphics{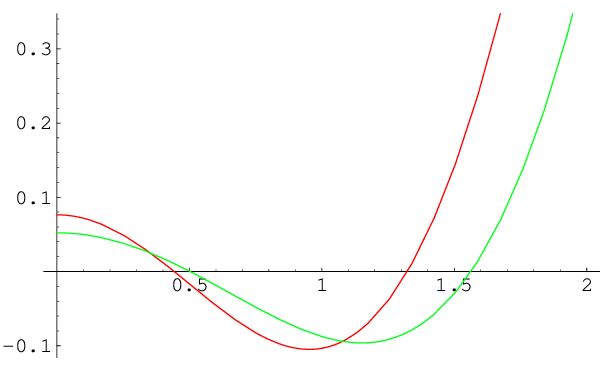}}}}
\vskip -6pt{
\caption{Graphs for $v_*(x)$ for $n=2$ from $\varepsilon$-expansion at \Oee\ and 
numerical solution with $v_*(0)=-V_*''(0)/d$ from Table 1 with  
$d=3.9$, $3.8$, $3.6$, $3.4$, $3.2$, $3.0$.}}
\end{figure}

Similarly we consider the multi-critical
fixed points obtained for $n=3$ and $n=4$, which correspond to $\vep$-expansions
for dimensions $d=3$ and $d=\frac{8}{3}$ respectively. 
When $n=3$ the results are
\be
k_3 =\frac{1}{5!^2} \, , \qquad
{a}_{2p}  = \frac{6-p}{3-p} \, \frac{9.2^{7-p}}{p!^2(6-p)!} \, , \quad
p=0,1,2,4,5 \, . 
\ee
Hence 
\be
{a}_{0} =\frac{16}{5}\, , \quad {a}_{2} =12\, , \quad
{a}_{4} =12\, , \quad {a}_8 =-\frac{1}{8}\, , \quad
{a}_{10} =-\frac{1}{800} \, .
\ee
{}From these, we can determine  from \eqref{eq:a2n2}
\be
{a}_{6} =\frac{173}{20} \, .
\ee
At the origin we then obtain
\be
v_*(0) =-\frac{1}{120}\vep\lt(1+\frac{1087}{120}\vep^2\rt) \, , \qquad
v''_*(0) =\frac{1}{10}\vep\lt(1+\frac{389}{40}\vep\rt) \, ,
\ee
and hence
\be
V_*''(0) = \frac{1}{40}\vep\lt(1+\frac{349}{40}\vep\rt) \, .
\label{Vpp3}
\ee
It is easy to verify that $V_*''(0)/V_*(0) = -d$ to this order again.
In Table~\ref{comparison2} the result \eqref{Vpp3} is compared
with the numerical results of \cite{HarveyFros} in this case and in 
Figure 2 a graphical comparison of $\vep$-expansion and numerical solutions 
is made.
\begin{table}[!h]
\begin{tabular}{|r|r|r|} \hline
$d$& Numerical& $\vep$-expansion at \Oee\\
\hline3&0~~&$0\qquad$\\
2.9&0.005~~&$0.005\qquad$\\
2.8&0.015~~&$0.014\qquad$\\
2.7&0.032~~&$0.027\qquad$\\
2.6&0.062~~&$0.045\qquad$\\
2.5&0.108~~&$0.067\qquad$\\
\hline
\end{tabular}\caption{Comparison of ERG numerical and analytical results for
$V_*''(0)$ for the $n=3$ fixed point.}\label{comparison2}
\end{table}
\begin{figure}[h]
\hskip -1.2cm{
\scalebox{0.5}{\includegraphics{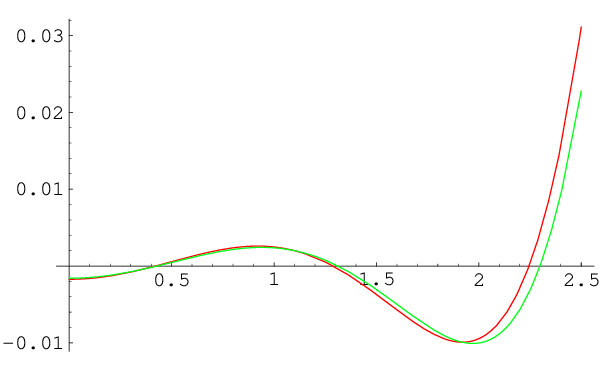}}
\scalebox{0.5}{\includegraphics{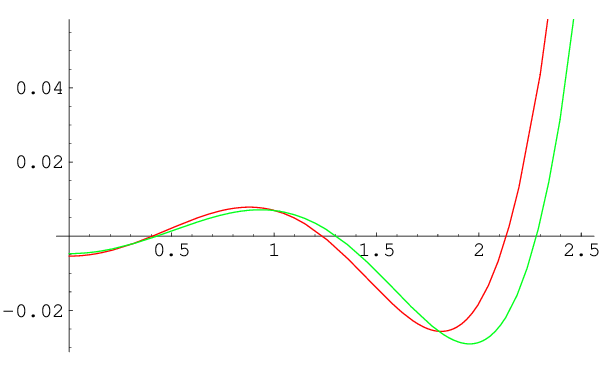}}
\scalebox{0.5}{\includegraphics{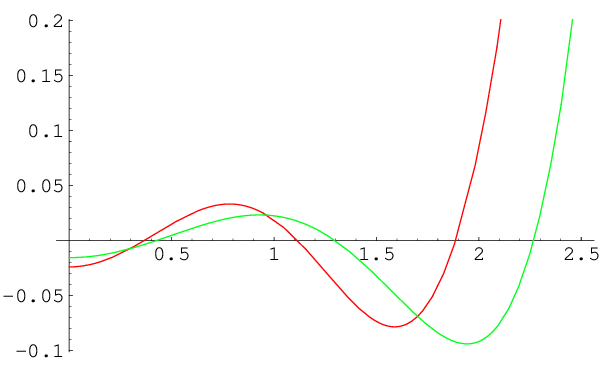}}}
\caption{$n=3$ $v_*(x)$ from $\varepsilon$-expansion at \Oee\
and  numerical solution for $d=2.9$, $2.8$ and $2.6$.}
\end{figure}

\begin{figure}[!h]
\hskip -1.2cm{
\scalebox{0.5}{\includegraphics{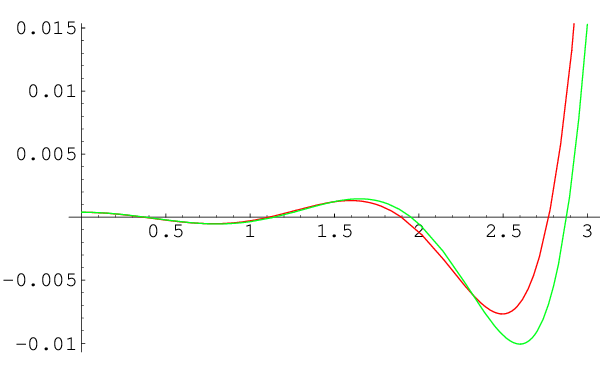}}
\scalebox{0.5}{\includegraphics{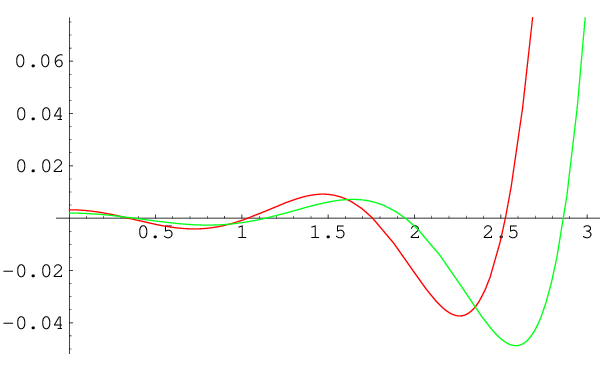}}
\scalebox{0.5}{\includegraphics{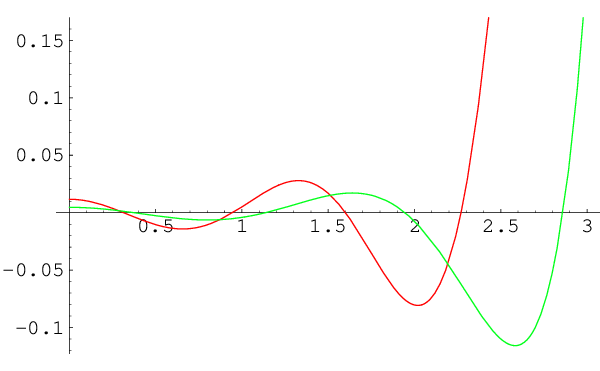}}}
\caption{$n=4$ $v_*(x)$ from $\varepsilon$-expansion at \Oee\
and  numerical solution for  $d=2.6$, $2.5$ and $2.4$.}\label{fig3}
\end{figure}

Finally, we consider the $n=4$ case. Following the same programme as
above, the coefficients in the expansion are
\begin{align}
{a}_{0} = & \frac{432}{35}\, , \quad {a}_{2} = \frac{288}{5}\, , \quad
{a}_{4} = \frac{324}{5} \, , \quad {a}_6 = 36\, , \nn \\
{a}_{10} =& -\frac{27}{100} \, , \quad {a}_{12} =-\frac{3}{800} \, , \quad
{a}_{14} =-\frac{1}{39200} \, , 
\end{align}
and
\be
a_8 = \frac{23904}{875} \, .
\ee
The fixed point solution at order $\vep^2$ then determines
\be
V_*''(0) = -\frac{3}{560}\vep\lt(1+\frac{99441}{3500}\vep\rt) \, .
\ee
Table~\ref{comparison3} compares this result with~\cite{HarveyFros} and
with some corresponding graphs exhibited in Figure \ref{fig3}.
\begin{table}[h]
\begin{tabular}{|r|r|r|} \hline
$d$&Numerical&$\vep$-expansion at \Oe\\
\hline$\frac{8}{3}$&0~~&$0\qquad$\\
2.6&$-0.001$~~&$-0.001\qquad$\\
2.5&$-0.008$~~&$-0.005\qquad$\\
2.4&$-0.028$~~&$-0.012\qquad$\\
2.3&$-0.082$~~&$-0.022\qquad$\\
2.2&$-0.226$~~&$-0.035\qquad$\\
\hline
\end{tabular}\caption{Comparison of ERG numerical and analytical results for
$V_*''(0)$ for the $n=4$ fixed point.}\label{comparison3}
\end{table}

\section{Critical Exponents}\label{sec:exponentsfirstorder}

Having computed $v_*(x)$ to order $\vep^2$ for the fixed points
below each critical dimension $d = 2n/(n-1)$, we now consider the RG flow
near these fixed points and compute certain critical exponents. In
the local potential approximation, the ERG flow is given
by~\eqref{eq:polch} and with the change of variables as in~\eqref{eq:redef}
we now have the following RG flow equation for $v(x,t)$
\begin{equation}
\frac{4}{d-2}\big (\dot{v}(x,t)-dv(x,t) \big )=
v''(x,t)-2x v'(x,t)-v'(x,t)^2 \, .
\end{equation}
In the neighbourhood of a fixed point 
\begin{equation}
v(x,t)=v_*(x)+e^{\lambda t}f(x) \, ,
\end{equation}
where $f(x)$ therefore satisfies the following linear eigenvalue equation
\begin{equation}
\frac{4}{d-2}(\lambda-d)=f''-2xf'-2v'_*f' \, .
\label{eq:linearized}
\end{equation}
This may be rewritten in the form
\begin{equation}
- D f(x) + v_*{\!}'(x) f'(x) = \hat{\lambda}f(x)  \, , \qquad
\hat{\lambda} = - \frac{2}{d-2}(\lambda -d) \, ,
\label{eq:linearized2}
\end{equation}
with $D$ the differential operator in \eqref{Dop}.

In the case of the Gaussian fixed point $v_*=0$, the eigenvalues are
$\hat{\lambda}_k=k$ and the associated eigenfunctions $f_k$ are just Hermite
polynomials, so that
\be
f_k(x)  =H_k(x) \, , \qquad
\lambda_k  = d-\half k(d-2) \, , \qquad k=0,1,2,\dots \, .
\ee
These correspond to the operators $\phi^k$ where $\phi$ has dimensions $\half(d-2)$.
For the high temperature fixed point then in \eqref{eq:linearized} 
$v_*'(x) = \frac{4}{d-2} x$ and the eigenfunctions are again Hermite
polynomials with rescaled variable and $\lambda_k  = d-\half k(d+2)$.

For non trivial critical points we require $v_*$ to be a non singular solution
of \eqref{eq:lpanonlin1.5} 
For dimension $d$ as in \eqref{dee} we may then consider 
a perturbation expansion in $\vep$, so that
\begin{align}
f_k(x)& = H_k(x)+ {\rm O}(\vep) \, , \nn\\
\hat{\lambda}_k & =
k+\vep\hat{\lambda}_k{\!}^{(1)} +\vep^2\hat{\lambda}_k {\!}^{(2)} +
{\rm O}(\vep^3) \, .
\label{eq:eigenlpa} 
\end{align}
To \Oe\ it is sufficient to take in \eqref{eq:linearized2}
\begin{equation}
v_*(x)=\ck_n\vep H_{2n}(x)\, .
\label{first}
\end{equation}
To extract $\hat{\lambda}_k{\!}^{(1)}$ we use standard first-order
perturbation theory using the basis of eigenfunctions of $ D$,
\begin{align}
N_k\hat{\lambda}_k{\!}^{(1)} & =\ck_n \int_{-\infty}^{\infty}\!\!\!\! \rmd x \, 
e^{-x^2}H_k(x)H_{2n}{\!}'(x)H_k{\!}'(x) \nn\\
& =-\half \ck_n\int_{-\infty}^{\infty}\!\!\!\! \rmd x \, H_k(x)^2
\frac{\rmd}{\rmd x}\lt(e^{-x^2}H_{2n} {\!}'(x) \rt)\nn\\
& =2n\ck_n\int_{-\infty}^{\infty}\!\!\!\! \rmd x \, e^{-x^2}H_k(x)^2H_{2n}(x)\nn\\
& =2n\ck_n \, G_{kk\,2n} \, . 
\end{align}
This gives
\begin{equation}
\hat{\lambda}_k{\!}^{(1)}=2(n-1)^2\frac{n!}{(2n)!}\frac{k!}{(k-n)!}\, ,
\label{eq:pert1}
\end{equation}
and in terms of the exponents $\lambda_k$, we have
\begin{align}
\lambda_k& =d-\half (d-2) \hat{\lambda}_k\nn\\
& =\frac{2n-k}{n-1}+\vep\lt(\frac{1}{2}k-1-2(n-1)
\frac{n!}{(2n)!}\frac{k!}{(k-n)!}\rt) + {\rm O}(\vep^2) \, .
\label{eq:exponents}
\end{align}
The results \eqref{eq:pert1} or equivalently \eqref{eq:exponents}  were found in 
the beginning of the RG analysis of critical points in \cite{Nicoll}, 
using an approximation to the Wegner-Houghton RG equation, see also \cite{Phase}, 
and (up to misprints) in~\cite{Haagensen}, using the LPA for 
the Wegner-Houghton equation. They are identical with the perturbative result
\eqref{etaone}.

For higher order calculations it is convenient to modify the eigenvalue equation in 
\eqref{eq:linearized2} by considering the transformed differential operator
\be
\Delta  =
e^{-v_*(x)}\Big ( - D + v_*{\!}'(x) \frac{\rmd}{\rmd x} \Big )e^{v_*(x)}
= -  D +\frac{2d}{d-2}\,  v_*(x)  \, , \\
\label{newop}
\ee
using that $v_*(x)$ satisfies the fixed point equation~\eqref{eq:lpanonlin1.5}. 
It is obvious that the eigenvalue equation
\be
\Delta {\hat f} = {\hat \lambda} {\hat f} \, ,
\label{Deig}
\ee
is equivalent to \eqref{eq:linearized2} and furthermore the operator $\Delta$ is 
hermitian\footnote{We may also note
$e^{-\frac{1}{2}x^2}\Delta \, e^{\frac{1}{2}x^2}$ is a Scr\"odinger operator
with potential $U(x)=\half x^2 - \half + \frac{2d}{d-2}\,  v_*(x)$. Asymptotically
$U(x) \sim \half \frac{(d+2)^2}{(d-2)^2}\, x^2$.}
with respect to the measure $\rmd x \, e^{-x^2}$.

To \Oe\ it is easy to see that, with $v_*$ given by \eqref{first} and using
the expansion \eqref{eq:expansion1},
that the eigenvalues are the same as \eqref{eq:pert1}. To \Oee\ and using
\eqref{eq:fp} second-order perturbation theory gives
\begin{align}
\hat{\lambda}_k{\!}^{(2)}&
=(n-1)^2\frac{\ck_n}{N_k}G_{2n\,kk}+\frac{2n\ck_n}{N_k}\sum_{p=0}^{2n-1}
a_{2p}G_{2p\,kk} -(2n\ck_n)^2\sum_{m\neq
k}\frac{1}{N_mN_k}\frac{G_{2n\,mk}^{\,2}}{m-k} \, ,
\end{align}
where the first two terms arise from the \Oee\ terms in
the operator itself, and the final term is the usual second order perturbation
expression for a perturbative potential $2n\ck_n \vep H_{2n}$. 
Substituting the expressions for $a_{2n}$
and $a_{2p} (p\neq n)$,~\eqref{eq:a2n2} and~\eqref{eq:a2p}
respectively, the expression above becomes
\begin{align}
\hat{\lambda}_k{\!}^{(2)}& =\lt((n-1)^2+n(n-1)\rt)
\frac{\ck_n}{N_k}G_{2n\,kk}-\frac{4n}{(n-1)^2}\frac{\ck_n}{N_{2n}}\sum_{p\neq n}
\frac{p(2n-p)}{2(n-p)}\frac{G_{2n\,2n\,2p}^{\,2}}{N_{2p}}\nn\\
&\ \ +2n\ck_n^2\sum_{p\neq
n}\frac{2n-p}{2(n-p)}\frac{G_{2p\,kk}G_{2n\,2n\,2p}}{N_{2p}N_k}-
(2n\ck_n)^2\sum_{m\neq k}\frac{1}{N_mN_k}\frac{G_{2n\,mk}^{\,2}}{m-k} \, .
\end{align}
Finally, substituting for $N_m$ from~\eqref{eq:Nm}, $G_{klm}$
from~\eqref{eq:Gklm} and $\ck_n$ from~\eqref{eq:kn}, we have
\begin{align}
\hat{\lambda}_k{\!}^{(2)}&=
(n-1)^3(2n-1)\frac{1}{n}\frac{n!k!}{(2n)!(k-n)!}\nn\\
&\ \ -2(n-1)^4\frac{1}{n^2}\frac{n!^7}{(2n)!^2} \frac{k!}{(k-n)!}
\sum_{p\neq n}\frac{p(2n-p)}{n-p}\frac{(2p)!}{p!^4(2n-p)!^2}\nn\\
&\ \ + (n-1)^4\frac{1}{n}\frac{n!^6}{(2n)!^2}k!\sum_{p\neq
n}\frac{2n-p}{n-p}\frac{(2p)!}{p!^4(k-p)!(2n-p)!}\nn\\
&\ \ - 2(n-1)^4\frac{n!^6}{(2n)!^2}k!\sum_{s\neq
n}\frac{1}{n-s}\frac{(2n+k-2s)!}{(k-s)!^2s!^2(2n-s)!^2} \, ,
\label{eq:pert2}
\end{align}
where $2s=2n+k-m$. While this expression is somewhat
complicated, for any particular choices of $n$ it can be simplified
to a polynomial in $k$ of order $2n-1$. Here we give the
results for the first two fixed points
\begin{subequations}
\begin{align}
\hat{\lambda}_{k}{\!}^{(2)}\big |_{n=2}=&-\frac{1}{12}k(k-1)(k-4) \, ,
\label{eq:pert2b}\\
\hat{\lambda}_{k}{\!} ^{(2)}\big |_{n=3}=&-\frac{1}{600}k(k-1)
(13k^3-17k^2-424k+800) \, .
\label{eq:pert2a}
\end{align}
\end{subequations}
A similar, but not identical result to~\eqref{eq:pert2b} 
was obtained in~\cite{Haagensen}, again using the LPA for
the Wegner-Houghton RG equation rather than the Polchinski equation.

Having computed the eigenvalues $\hat{\lambda}$ to \Oee,
we may now extend~\eqref{eq:exponents} to calculate the
corresponding critical exponents to \Oee. The exponents
$\lambda_k$ are given in terms of the eigenvalues as
\begin{align}
\lambda_k & =\frac{2n-k}{n-1}+\vep\lt(\frac{k}{2}-1-2(n-1)
\frac{n!}{(2n)!}\frac{k!}{(k-n)!}\rt)\nn\\
&\ \
+\vep^2\lt((n-1)^2\frac{n!}{(2n)!}\frac{k!}{(k-n)!}-
\frac{1}{n-1}\hat{\lambda}_k{\!}^{(2)}\rt) + {\rm O}(\vep^3) \, .
\end{align}
This gives
\begin{align}
\lambda_k \big |_{n=2}  = {}& 4-k + \vep \big ( \half k - 1 -
\textstyle{\frac{1}{6}} k(k-1) \big ) + \vep^2 \, 
\textstyle{\frac{1}{12}}k(k-1)(k-3) + {\rm O}(\vep^3) \, , \nn \\
\lambda_k \big |_{n=3} = {}& \half (6-k) + \vep \big ( \half k - 1 -
\textstyle{\frac{1}{30}} k(k-1)(k-2) \big ) \nn \\
&{} + \vep^2 \, \textstyle{\frac{1}{1200}}\, k(k-1)(k-5)(13k^2+48k-144) 
+ {\rm O}(\vep^3) \, .
\end{align}

\subsection{Exact Exponents}

The results \eqref{eq:pert1} and \eqref{eq:pert2b}, \eqref{eq:pert2a} show
that ${\hat \lambda}_k =0$ for $k=0,1$ to \Oee, at least for $n=2,3$. 
For general $n$ from \eqref{eq:pert2}
\begin{align}
\hat{\lambda}_1{\!}^{(2)}&
=(n-1)^4\frac{1}{n}\frac{n!^6}{(2n)!^2}
\sum_{p=0}^{1}\frac{2n-p}{n-p}\frac{(2p)!}{p!^4(1-p)!(2n-p)!}
\nn\\& \ \ \ -2(n-1)^4\frac{n!^6}{(2n)!^2}
\sum_{s=0}^1\frac{1}{n-s}\frac{(2n+1-2s)!}{(1-s)!^2s!^2(2n-s)!^2}\nn\\
& =2(n-1)^4\frac{n!^6}{(2n)!^2}\frac{1}{n(n-1)(2n)!}
\lt((n-1)(-2n)+2n(2n-1)-2n^2\rt)\nn\\
& =0 \, .
\label{eq:pert3}
\end{align}

These results follow in general
since it is possible to find exact eigenfunctions for  $k=0,1$ in
\eqref{eq:linearized2}.
Firstly we have the trivial case,
\begin{equation}
f_0=1 \, , \qquad \hat{\lambda}_0=0 \, .
\end{equation}
For $k=1$ we also have
\begin{equation}
f_1(x) =v'_*(x)-\frac{4x}{d-2}\, ,\qquad \hat{\lambda}_1=1 \, .
\end{equation}
To verify this we first obtain from \eqref{eq:lpanonlin1.5}
\be
Dv_* {\!}' + \frac{d+2}{d-2}\, v_* {\!}' =  v_* {\!}' v_* {\!}'' \, ,
\label{vprim}
\ee
since then
\be
D f_1(x) - v_* {\!}'(x) f_1(x) = - v_* {\!}'(x) + \frac{4x}{d-2} = - f_1(x)\, .
\ee
The vanishing of $\hat{\lambda}_1$ reflects that there is no anomalous
dimension for the field $\phi$ in the LPA.

Additionally from \eqref{vprim} we have another exact eigenfunction and eigenvalue,
\begin{equation}
f_{2n-1}(x) =v'_*(x) \, ,\qquad
\hat{\lambda}_{2n-1}=\frac{d+2}{d-2} \, .
\label{eq:exactexp}
\end{equation}
Expanding the exact eigenvalue $\hat{\lambda}_{2n-1}$ in
powers of $\vep$ gives
\begin{equation}
\hat{\lambda}_{2n-1} =2n-1+(n-1)^2\vep +\half (n-1)^3\vep^2 +\dots \, ,
\end{equation}
which is consistent with~\eqref{eq:pert1} since
\be
\hat{\lambda}^{(1)}_{2n-1} =(n-1)^2 \, .
\ee
It is also in accord with the second-order
result,~\eqref{eq:pert2}. Relabelling $s=2n-p$,
\begin{align}
\hat{\lambda}_{2n-1}^{(2)}&
=(n-1)^3(2n-1)\frac{1}{n}\frac{n!(2n-1)!}{(2n)!(n-1)!}\nn\\
& \ \ \ -2(n-1)^4\frac{1}{n^2}\frac{n!^7}{(2n)!^2}
\frac{(2n-1)!}{(n-1)!}
\sum_{p\neq n}\frac{p(2n-p)}{n-p}\frac{(2p)!}{p!^4(2n-p)!^2}\nn\\
& \ \ \ + (n-1)^4\frac{1}{n}\frac{n!^6}{(2n)!^2}(2n-1)!\sum_{p\neq
n}\frac{2n-p}{n-p}\frac{(2p)!}{p!^4(2n-1-p)!(2n-p)!}\nn\\
& \ \ \ + 2(n-1)^4\frac{n!^6}{(2n)!^2}(2n-1)!\sum_{p\neq
n}\frac{1}{n-p}\frac{(2p-1)!}{(p-1)!^2p!^2(2n-p)!^2}\nn\\
& =\frac{1}{2}(n-1)^3+\frac{(n-1)^4}{2n^2}\frac{n!^6}{(2n)!}
\sum_{p=0}^{2n}(4n-3p)\frac{(2p)!}{p!^4(2n-p)!^2} \, .
\label{sumtwo}
\end{align}
Provided the final sum is identically zero for all $n$ this
agrees with the exact result. We demonstrate that the sum
vanishes in Appendix~\ref{sec:vanishingsum}. In this case the relevant
operator $\phi^{2n-1} \propto \pr^2 \phi$ by the equations of motion and
has dimension $\half(d+2)$.

\section{Beyond the Local Potential Approximation}\label{sec:2od}

Although the LPA captures the essential features of the fixed point
structure in the space of couplings for scalar theories it neglects 
the momentum dependence of vertices and is therefore limited in terms of 
calculating critical exponents quantitatively, the anomalous dimension
$\eta$ of the scalar field is undetermined and set to zero. Although at the
Wilson-Fisher fixed point $\eta$ is small and in the $\vep$-expansion $\eta={\rm O}
(\vep^2)$ it is of course necessary to take $\eta$ into account in more systematic
treatments. To this end a natural extension of the LPA
is to assume a solution of the exact RG flow equations which is expressible
in terms of a local functional of the fields and their derivatives, the
derivative expansion \cite{Morris2}, for a recent discussion see \cite{Deriv}. 
In the derivative expansion there is a 
necessary dependence on the form of the cut off function but in application to the 
Polchinski RG equation, with terms quadratic in derivatives, there are two
constants,  see \cite{Ball,Com}.

At the second order in the derivative expansion, the Polchinski equation may 
be reduced \cite{Ball} to a following pair of coupled ODEs, extending \eqref{eq:polch}, 
for a potential $V(\phi,t)$ and also the coefficient $Z(\phi,t)$ of 
$\half (\pr \phi)^2$ in the derivative expansion,
\begin{subequations}
\begin{align}
\dot{V}& =V''-\half(d-2+\eta)\phi V'+dV-V'^2+AZ \, , \label{couplea} \\
\dot{Z}& =Z''-\half(d-2+\eta)\phi Z'-2V'Z'-4V''Z-\eta Z-\eta
+2BV''^2 \,  ,
\label{coupleb}
\end{align}
\end{subequations}
where $A$ and $B$ are the two cut off dependent constants which cannot be eliminated
by any rescaling (essentially the same equations were obtained from the Wilson
RG equations in \cite{Golner}). 

As earlier it is convenient in our discussion to introduce a rescaled variable
\be
x=\half(d-2+\eta)^{\frac{1}{2}}\phi \, , \qquad
v(x,t) =V(\phi,t)\, ,\quad
z(x,t)  =Z(\phi,t),
\ee
so that the coupled equations \eqref{couplea}, \eqref{coupleb} become
\begin{subequations}
\begin{align}
\frac{2}{d-2+\eta}\, \dot{v}& = (D+K)v- \half \, v'^2+ \tilde{A}z \, , 
\label{eq:2orga}  \\
\frac{2}{d-2+\eta}\, \dot{z}& = (D-L)z  -v'z'- 2v''z - L + \half \tilde{B}\, v''^2 \, ,
\label{eq:2orgb}
\end{align}
\end{subequations}
where 
\be
K =  \frac{2d}{d-2+\eta} \, , \quad L= \frac{2\eta}{d-2+\eta} \, , \qquad
\tilde{A}=\frac{2}{d-2+\eta}A \, , \quad
\tilde{B}=\half (d-2+\eta)B \, .
\label{KLAB} 
\ee
At a fixed point, $v(x,t)\to v_*(x), \, z(x,t) \to z_*(x)$ which satisfy the equations 
\begin{subequations}
\begin{align}
(D  + K )v_*& = \half\,  v_*{\!}'^2 - \tilde{A}z_* \, , \quad
\label{eq:2ov}\\
(D - L ) z_* - v_*{\!} 'z_*{\!}' - 2v_*{\!} ''z_*&
= L  -  \half \tilde{B} \, v_*{\!}''^2 \,  .
\label{eq:2oz}
\end{align}
\end{subequations}
Assuming \eqref{dee} then in an $\vep$-expansion, assuming $A,B$ are $O(1)$,
a consistent solution is obtained by requiring $v_*(x)= {\rm O}(\vep)$ and 
$\eta, z_*(x)= {\rm O}(\vep^2)$. To lowest order then $v_* \to \vep \ck_n H_{2n}$
as in \eqref{vnn}. With this result then \eqref{eq:2oz} becomes
\be
D z_* = (n-1)\eta - \vep^2 \half \tilde{B}\ck_n{\!}^2 H_{2n}{\!\!}''^2 
+ {\rm O}(\vep^3) \, .
\label{eq:2oz2}
\ee
This determines $\eta=\vep^2 \, \eta^{(2)}+ {\rm O}(\vep^3)$ since the left hand side
of \eqref{eq:2oz2} is  orthogonal to 1 so that
\be
(n-1)\eta^{(2)} N_0  = \half \tilde{B}\ck_n {\!}^2 
\int^{\infty}_{-\infty} \!\!\!\! \rmd x \,e^{-x^2} H_{2n}{\!\!}''^2 \, ,
\ee
which implies
\be
\eta^{(2)}  = 2^{2n+2} \frac{n(2n-1)}{n-1} (2n)! \, \tilde{B} \ck_n{\!}^2 \, .
\label{eta2}
\ee
We consider this further later but first we may then use \eqref{eq:2oz2}
to determine $z_*(x)$ to lowest order by using the expansion
\be
z_*=\vep^2  \ck_n{\!}^2 \sum_p b_{2p}H_{2p}  + {\rm O}(\vep^3)\, .
\label{eq:zexp}
\ee
Since $DH_{2p} = -4pH_{2p}$ and with $H_{2n}{\!\!}''=8n(2n-1)H_{2n-2}$
we may obtain
\begin{align}
b_{2p}  = {}& 16\tilde{B}\, n^2(2n-1)^2 \frac{G_{2p\,2n-2\,2n-2}}{pN_{2p}} \nn \\
={}&  2^{2n-p} \tilde{B} \frac{(2n)!^2}{p(2n-2-p)!p!^2} \, , \quad
p=1,\dots,2n-2 \, .
\label{eq:b2p}
\end{align}
We note that $b_0$ is not constrained by \eqref{eq:2oz2}. Nevertheless $b_0$ may
be determined by imposing $z_*(0)=0$ \cite{Ball}. This ensures that \eqref{eq:2ov},
\eqref{eq:2oz} have well defined solutions for all $x$ only for a specific choice
of $\eta$ and $v_*(0)$.

With these results \eqref{eq:2ov} becomes
\be
(D+2n) v_* + (n-1)^2 \vep v_* = \vep^2 \ck_n{\!}^2 \bigg ( H_{2n}{\!\!}'^2 -
{\tilde A} \sum_{p=0}^{2n-2}  b_{2p} H_{2p} \bigg ) + {\rm O}(\vep^3) \, ,
\label{vAB}
\ee
where $v_*$ is expressible as in \eqref{eq:fp} and $b_{2p}$ is given by
\eqref{eq:b2p}. Just as in \eqref{nonl} and \eqref{eq:kn} this determines
normalisation coefficient $\ck_n$,
\be
(n-1)^2\ck_nN_{2n} = \ck_n{\!}^2 \big ( n  G_{2n\,2n\,2n} - {\tilde A} N_{2n} 
b_{2n} \big ) \quad
\Rightarrow \quad \ck_n \Big ( 1 - \frac{n-1}{n}\, {\tilde A}{\tilde B} \Big )
=\frac{(n-1)^2}{2^nn} \,\frac{n!^3}{(2n)!^2} \, . 
\label{eq:kn2}
\ee
Furthermore in \eqref{eq:fp} the expansion coefficients are also determined
by \eqref{vAB}
\begin{align}
a_{2p}& =  \ck_n\frac {1}{2(n-p)}\
\Big (  (2n-p)\,  \frac{G_{2p\,2n\,2n}}{N_{2p}} - {\tilde A}\, b_{2p} \Big ) \nn\\
& =  \ck_n \,  \frac{2^{2n-p}}{2(n-p)} \frac{(2n)!^2}{p!^2(2n-1-p)!} 
\Big (1 - \frac{2n-1-p}{p} \,{\tilde A}{\tilde B} \Big )  \, , 
\quad  p\ne n\, .
\label{a2p}
\end{align}
except for $a_{2n}$. To obtain this it is necessary to extend \eqref{vAB}
to ${\rm O}(\vep^3)$.

Using the result \eqref{eq:kn2} the expression \eqref{eta2} becomes
\begin{equation}
\eta^{(2)} =4(n-1)^2\frac{n!^6}{(2n!)^3} \,  \frac{n(n-1)(2n-1){\tilde B}}
{\big (n -(n-1){\tilde A}{\tilde B} \big )^2} \, . 
\end{equation}
The perturbative result \eqref{etaone} requires that the final factor,
depending on $A,B$, should be one.

\section{Critical Exponents with the Derivative Expansion}
\label{sec:exponentssecondorder}

We now consider small departures from the fixed point of the form
\begin{align}
v(x,t)& =v_*(x)+e^{\lambda t}f(x)\, , \nn\\
z(x,t)& =z_*(x)+e^{\lambda t}g(x) \, .
\end{align}
Substituting this form into \eqref{eq:2orga}, \eqref{eq:2orgb} 
we find the eigenvalue equations
\begin{subequations}
\begin{align}
-(D+K) f + v_*{\!}' f' - \tilde{A}\, g ={}&  \hat{\lambda}f \, , 
\label{eone} \\
-(D-L)g +v_*{\!}'  g' + 2v_*{\!}''  g + z_*{\!}' f'
+ 2 z_* f'' -  \tilde{B}\, v_*{\!} '' f''  = {}&  \hat{\lambda}f \, ,
\label{etwo}
\end{align}
\end{subequations}
with $D$ as in \eqref{Dop} and
where $\lambda$ is related to $\hat{\lambda}$ according to
(this differs from \eqref{eq:linearized2}).
\begin{equation}
\lambda = - \half (d-2+\eta) \hat{\lambda} \, .
\label{eek}
\end{equation}
Defining the vector
\begin{equation}
F=\lt(\begin{array}{c}f \\ g \\ \end{array}\rt) \, ,
\end{equation}
then \eqref{eone},  \eqref{etwo} may be written more compactly in matrix form  as
\be
\Delta F  =\hat{\lambda} F \, .
\label{mat}
\ee 
The operator $\Delta$ is not hermitian so for later convenience we also
consider the dual equation defining the dual eigenvectors
\be
{\tilde \Delta} \tilde{F} =\hat{\lambda}\tilde{F} \, ,
\ee
where ${\tilde \Delta}$ is the adjoint of $\Delta$ with a scalar product
defined  $\int {\rm d} x \, e^{-x^2}\, F_1{\!}^T F_2$.

Just as in Section \ref{sec:exponentsfirstorder}
there are exact eigenfunctions and eigenvalues \cite{Ball}
\begin{subequations}
\begin{align}
F = {}& \lt(\begin{array}{c}1 \\ 0 \\ \end{array}\rt) \, ,  \qquad\qquad\qquad\quad
\ \  \hat{\lambda}= - K \, , \qquad \ \ \lambda =   d \, ,  \label{exact1} \\
F = {}&  \lt(\begin{array}{c} v_*{\!}'- (K-2)x \\ z_*{\!}'  \\ \end{array}\rt) \, , 
\qquad \hat{\lambda}  =   - K+ 1  \, , \quad 
\lambda =   \half(d+2-\eta) \, ,  \label{exact2} \\
F ={}&  \lt(\begin{array}{c}v_*{\!}' \\ z_*{\!}' \\ \end{array}\rt) \, ,  
\qquad\qquad\qquad\quad
\, \hat{\lambda} =  - 1 \, , \qquad \quad \, \lambda = \half (d-2+\eta) \, .
\label{exact3}
\end{align}
\end{subequations}
The first corresponds to the identity operator and at the $n$-th multicritical 
point the second and third to the operators $\phi$ and $\phi^{2n-1}$.

In the $\vep$-expansion
\begin{align}
\Delta = \Delta ^{(0)}+\vep \Delta^{(1)}+\dots \, ,
\label{delexp}
\end{align}
and correspondingly in \eqref{mat}
\begin{equation}
\hat{\lambda}=\hat{\lambda}^{(0)}+\vep\hat{\lambda}^{(1)}+\dots \, , \qquad
F = F^{(0)} + \vep F^{(1)} + \dots \, , 
\end{equation}
and similarly for $\tilde{F}$.
At zeroth order in $\vep$, 
\begin{equation}
\Delta^{(0)}=\lt(\begin{array}{cc}-D -2n & -\tilde{A} \\
0 & -D  \\ \end{array}\rt) \, .
\label{del0}
\end{equation}
There are two sets of eigenfunctions, which are easily obtained
\be
F^{(0)}_{k,1}=\lt(\begin{array}{c}1 \\ 0\\ \end{array}\rt) H_k \, , \qquad  
\tilde{F}^{(0)}_{k,1}=\lt(\begin{array}{c}1 \\
\frac{\tilde{A}}{2n}\\\end{array}\rt)H_k \, , \qquad k=0,1,2,\dots \, ,
\ee
and 
\be
F^{(0)}_{k,2}=\lt(\begin{array}{c}-\frac{\tilde{A}}{2n} \\
1\\ \end{array}\rt) H_{k-2n} \, , \qquad 
\tilde{F}^{(0)}_{k,2}=\lt(\begin{array}{c}0\\
1 \\\end{array}\rt)H_{k-2n} \, , \qquad k=2n,2n+1,\dots \, .
\ee
The eigenvalues in both cases are
\begin{align}
\hat{\lambda}_{k,1}^{(0)} = \hat{\lambda}_{k,2}^{(0)} = k-2n \, ,
\label{eq:eigen2o}
\end{align}
which are thus two-fold degenerate for $k\ge 2n$.

In general in terms of the perturbative anomalous dimensions at the fixed
point discussed in Section \ref{sec:pertex}
\be
\lambda_{k,1} = \frac{2n-k}{n-1} - \omega_{k,1} \, , \qquad
\lambda_{k,2} = \frac{2n-k}{n-1} - \omega_{k,2} \, .
\label{lom}
\ee

\subsection{Eigenvalues at \Oe}

We now use first-order perturbation theory to compute
$\hat{\lambda}^{(1)}_{k,1}$ and $\hat{\lambda}^{(1)}_{k,2}$. In the expansion
\eqref{delexp} at \Oe\ 
\begin{equation}
\Delta^{(1)}=\lt(\begin{array}{cc}-(n-1)^2+\ck_nH_{2n}{\!}'\frac{\rmd}{\rmd x} & 0 \\
- \tilde{B}\, \ck_nH_{2n}{\!\!}'' \frac{\rmd^2}{\rmd x^2} & 
\ck_nH_{2n}{\!\!}'\frac{\rmd}{\rmd x}+2\ck_nH_{2n}{\!\!}'' \\
\end{array}\rt)\, .
\label{del1}
\end{equation}
The \Oe\ results are determined in terms of the $2\times 2$ matrix ${\cal M}_k$ 
defined by
\begin{equation}
{\cal M}_{k,ij}  =
\int_{-\infty}^{\infty}\!\!\!\! \rmd x \,
e^{-x^2} \tilde{F}^{(0)\, T}_{k,i}\,  \Delta^{(1)} F^{(0)}_{k,j}\, .
\end{equation}
For $k<2n$ there is a single eigenfunction so that the perturbation theory result is
just
\begin{align}
N_k\hat{\lambda}^{(1)}_{k,1} & = {\cal M}_{k,11} \nn\\
& = -(n-1)^2N_{k}+\ck_n\int_{-\infty}^{\infty}\!\!\!\! \rmd x \,
e^{-x^2}H_{k}H_{2n}{\!\!}' H_{k}{\!}' -\ck_n\frac{1}{2n}\tilde{A}\tilde{B}
\int_{-\infty}^{\infty}\!\!\!\! \rmd x \, e^{-x^2}H_{k}H_{2n}{\!\!}'' H_{k}{\!}'' \, .
\label{M11}
\end{align}
Hence, using the formulae for $N_k$, $G_{klm}$ in
\eqref{eq:Nm}, \eqref{eq:Gklm}  we obtain
\begin{align}
\hat{\lambda}^{(1)}_{k,1} ={}& -(n-1)^2 + \ck_n \, 2^{n+1} 
\frac{(2n)!k!}{n!^2 (k-n)!}\big (n-(n-1)\tilde{A}\tilde{B}(n-1)\big )\nn \\
= {}& - (n-1)^2 + 2(n-1)^2\frac{n!}{(2n)!}\frac{k!}{(k-n)!}\, ,
\label{eq:hatlam1}
\end{align}
where $\ck_n$ is determined by \eqref{eq:kn2} so that the dependence on 
$\tilde{A},\tilde{B}$ disappears. This result is essentially identical
to \eqref{eq:pert1} and is in accord with the exact results \eqref{exact1}, 
\eqref{exact2}, \eqref{exact3}  which correspond to $k=0,1,2n-1$.

For $k\ge 2n$ there is a two-fold degeneracy, and so we use
degenerate perturbation theory. In particular, for non-trivial
first-order eigenfunctions we require that the first-order
perturbations to the eigenvalues solve the following characteristic
equation
\begin{align}
\det \big [ \lambda^{(1)}_k{\cal N}_k - {\cal M}_k  \big ]=0 \, ,
\label{eq:degpertdet}
\end{align}
where the ${\cal N}_k$ is the diagonal matrix
\begin{equation}
{\cal N}_k=\lt(\begin{array}{cc}N_{k} & 0  \\0 & N_{k-2n} \\ 
\end{array}\rt) \, ,
\end{equation}
More explicitly, the elements
of $\cal M$ are given by \eqref{M11} and
\begin{subequations} 
\begin{align}
{\cal M}_{k,22}
& =\ck_n\int_{-\infty}^{\infty}\!\!\!\!  \rmd x \, 
e^{-x^2}H_{k-2n}^{\vphantom T} \lt(H'_{2n}H'_{k-2n}+2H''_{2n}H_{k-2n} + 
\frac{1}{2n}\tilde{A}\tilde{B} \, H''_{2n}H''_{k-2n} \rt ) \, , \\
{\cal M}_{k,21}
& = -\tilde{B}\, \ck_n\int_{-\infty}^{\infty} \!\!\!\!  \rmd x \,
e^{-x^2}H_{k-2n}^{\vphantom T} H''_{2n}H''_k,\\
{\cal M}_{k,12}
& = \ck_n\frac{1}{2n}\tilde{A}\int_{-\infty}^{\infty} \rmd x \,
e^{-x^2}H_k^{\vphantom T} \lt(2H''_{2n}H_{k-2n}^{\vphantom T} 
+\frac{1}{2n}\tilde{A}\tilde{B}H''_{2n}H''_{k-2n}\rt)\nn\\
& = 0 \, ,
\label{eq:b12}
\end{align}
\end{subequations}
where in \eqref{eq:b12} we use that $H_k$ is orthogonal to polynomials of degree $<k$.
Hence since the matrix ${\cal M}_k$ is lower triangular the eigenvalues solving 
\eqref{eq:degpertdet} are just $\hat{\lambda}^{(1)}_{1,k}$ as in \eqref{eq:hatlam1} and 
\begin{align}
\hat{\lambda}^{(1)}_{k,2} = {}& \frac{{\cal M}_{k,22}}{N_{k-2n}}
= \ck_n \, 2^{n+1}\frac{(2n)!(k-2n)!}{n!^2(k-3n+1)!}
\lt(n(k-n+1)+\tilde{A}\tilde{B}(n-1)(k-3n+1)\rt) \, .
\end{align}

{}From \eqref{eek} $\lambda = -(\frac{1}{n-1}- \half \vep) {\hat \lambda} 
+ {\rm O}(\vep^2)$ so to \Oe\ we then have
\begin{subequations}
\begin{align}
\lambda_{k,1} ={}& \frac{2n-k}{n-1}
+\vep\lt[\frac{1}{2}k-1-2(n-1)\frac{n! \,k!}{(2n)! \, (k-n)!} \rt] \, , 
\label{eq:splitting1}  \\
\lambda_{k,2} ={}& \frac{2n-k}{n-1} \nn \\
& \! {} +\vep\lt[\frac{1}{2}k-n- 2(n-1) \frac{n!\, (k-2n)!}{(2n)!\, (k-3n+1)!} 
\frac{n(k-n+1) + (n-1)\tilde{A}\tilde{B}(k-3n+1)} {n - (n-1) \tilde{A}\tilde{B} }\rt] \, .
\label{eq:splitting2}
\end{align}
\end{subequations}
The result \eqref{eq:splitting1} matches the perturbative result in
\eqref{expk} but \eqref{eq:splitting2} depends on $AB$ so cannot
agree with \eqref{expk2} in general.

\section{Expansion of Exact RG Equation}\label{Scaling}

An alternative approach, which nevertheless has many similarities although
some crucial differences to the derivative expansion, is to expand the full 
renormalisation group equation which contains linear and quadratic terms
in terms of translation invariant operators, or scaling fields, which are
exact eigen-solutions for its linearised part. This has been applied
in \cite{Phase} and \cite{Gol}, \cite{Rie} to the Wilson RG equation. Here we 
apply similar methods more simply to the Polchinski equation for the
effective action $S[\vphi,t]$ for a scalar field $\vphi$. This takes the
form, after convenient rescalings by the cut off scale $\Lambda$, $t=-\ln \Lambda$, 
\begin{align}
\frac{\pr}{\pr t} S = {} & \frac{1}{(2\pi)^d} \int \rmd^d p \, \big ( \half d + 1 - 
\half \eta + p \cdot \pr_p \big ) \tphi(p) \frac{\delta S}{\delta \tphi(p)}  \nn \\
&{} - \frac{1}{(2\pi)^d}\int \rmd^d p \, K'(p^2) \bigg ( \frac{\delta^2 S}
{\delta \tphi(p) \delta \tphi(-p)}
- \frac{\delta S}{\delta \tphi(p)}\, \frac{\delta S}{\delta \tphi(-p)} \bigg ) \nn \\
&{} - \half \eta\, \frac{1}{(2\pi)^d}\int \rmd^d p \, 
K(p^2)^{-1}p^2 \tphi(p)\tphi(-p) + C \, , 
\label{full}
\end{align}
where $K(p^2/\Lambda^2)$ is a cut off function and $\tphi$ is the Fourier 
transform of $\vphi$, $\delta \tphi(p)/\delta \tphi(p') = (2\pi)^d\delta^d(p-p')$.
$C$ is an additional constant independent of $\vphi$, in general it is irrelevant
and may be neglected\footnote{The standard derivation gives $C=\delta^d(0)
\int \rmd^d p \, K'(p^2) K(p^2)^{-1}p^2$ with $(2\pi)^d \delta^d(0)$ the overall
volume.}.
Apart from $K(0)=1$ and sufficient rapid fall off for large $p^2$ no restriction
on the cut off function $K(p^2)$ is imposed here.
The Gaussian fixed point corresponds to $S=0$ and also when the anomalous 
scale dimension for $\vphi$, $\eta=0$. The appearance of $\eta$ in \eqref{full}
arises by assuming that $\vphi$ varies with the cut off $\Lambda$ with an anomalous
dimension $\half \eta$. As will be apparent later the additional term proportional
to $\eta$ in \eqref{full} is necessary for consistent RG flow solutions. 
The derivative expansion is obtained
directly by approximating \eqref{full} by assuming $\tphi(p)$ is expanded 
as $(2\pi)^{d} (\phi -i  \pr \phi \cdot \pr_p ) \delta^d(p)$
and requiring  $S[\vphi,t] \to \int \rmd^d x \big ( V(\phi,t) + 
\half Z (\phi,t) (\pr \phi)^2 \big )$.

The starting point of the discussion in this Section requires solutions of 
\be
\big ( \D_1 + \D_2 \big ) \OO = \lambda \OO \, , 
\label{DDO}
\ee
for $\D_1 + \D_2$ the differential operator defined by the linear part of \eqref{full},
\begin{align}
\D_1 = {}& \frac{1}{(2\pi)^d}\int \rmd^d p \, 
\big ( \half d + 1  + p \cdot \pr_p \big ) 
\tphi(p) \, \frac{\delta}{\delta \tphi(p)} \, , \nn \\
\D_2 = {}& - \frac{1}{(2\pi)^d}\int \rmd^d p \, K'(p^2)  \frac{\delta^2}
{\delta \tphi(p) \delta \tphi(-p)} \, .
\label{D12}
\end{align}
The eigenvalue equation
\be
\D_1 {\hat \OO}  =  \lambda {\hat \OO} \, ,
\ee
is easily solved in terms of local translational invariant operators by
\be
{\hat \OO}  = \frac{1}{(2\pi)^{d(k-1)}}\int \prod_{i=1}^k \rmd^d p_i \, \tphi(p_i) \ 
\delta^d \big ({\textstyle \sum_i} p_i \big ) \, O(p_1,\dots, p_k) \, , \quad
\lambda= d  - k (\half d-1) - r \, , 
\label{hatO}
\ee
where $O(p_1,\dots, p_k)$ is a scalar symmetric homogeneous polynomial of degree $r$. 
We then obtain corresponding solutions of \eqref{DDO} in the form
\be
\OO = e^{-\Y} {\hat \OO}  \, ,
\label{YO}
\ee
where $\Y$ is defined by
\be
[ \D_1 , \Y ] = \D_2 \, ,
\label{DY}
\ee
ensuring that $\OO$ as defined by \eqref{YO} has the same eigenvalue as $\hat\OO$.
It is easy to solve \eqref{DY} giving
\be
\Y = \half \,  \frac{1}{(2\pi)^d}\int \rmd^d p \  G(p) \, \frac{\delta^2}
{\delta \tphi(p) \delta \tphi(-p)} \, , \qquad
G(p) = \frac{K(p^2)}{p^2} \, ,
\label{GY}
\ee
so that in \eqref{YO} $e^{-\Y}$ essentially generates normal ordering.

For comparison with the derivative expansion earlier we consider just operators with
$r=0,2$ for which a convenient basis is 
\begin{align}
\OO_{k0} = {}& e^{-\Y} \frac{1}{(2\pi)^{d(k-1)}}\int \prod_{i=1}^k \rmd^d p_i \, 
\tphi(p_i) \ \delta^d \big ( p_{(k)} \big ) \, , \qquad  
p_{(k)} = {\textstyle \sum_{i=1}^k}\, p_i \, , \nn \\
\OO_{k2} = {}& \frac{1}{k(k-1)} \, e^{-\Y}\,  \frac{1}{(2\pi)^{d(k-1)}}
\int \prod_{i=1}^k \rmd^d p_i \, \tphi(p_i) \ \delta^d \big ( p_{(k)} \big ) \, 
\sum_{i=1}^k p_k{\!}^2 \, , \quad k=2,3,\dots \, .
\end{align}
The result for $\OO_{k0}$ may be expressed in terms of Hermite polynomial as a 
consequence of the identity
\be
e^{-\frac{1}{4} \, \frac{\rmd^2}{\rmd x^2} } (2x)^n = H_n(x) \, .
\ee

The nonlinear term in \eqref{full} may be evaluated in terms operators $\OO$ and
$\OO'$, respectively of degree $l$ and $m$ in $\tphi$ and as given by \eqref{hatO} 
and \eqref{YO}, by considering \cite{Phase}
\begin{align}
&  e^{\Y} \frac{1}{(2\pi)^d} \int \rmd^d p \, K'(p^2) \,
\frac{\delta \OO}{\delta \tphi(p)} \, 
\frac{\delta \OO'}{\delta \tphi(-p)} \nn \\ 
&{} = \exp \bigg ( \frac{1}{(2\pi)^d} \! \int \! \rmd^d p \, G(p) \, \frac{\delta^2}
{\delta \tphi(p) \delta \tphi'(-p)} \bigg ) \, lm
\frac{1}{(2\pi)^{d(l+m-3)}}\! \int \prod_{i=1}^{l-1} \rmd^d p_i \, \tphi(p_i) 
\! \int \prod_{j=1}^{m-1} \rmd^d q_j \, \tphi'(q_j) \nn \\
\noalign{\vskip -4pt}
&{}\! \quad \times \delta^d\big ( p_{(l-1)} + q_{(m-1)}\big ) \ 
O\big (p_1,\dots,p_{l-1},-p_{(l-1)}\big ) \,
O'\big (q_1,\dots,q_{m-1},-q_{(m-1)}\big ) \Big |_{\tphi'=\tphi} \, 
K'\big (p_{(l-1)}{\!\!}^2 \big )\nn \\
&{} = \sum_{n\ge 0} \frac{ l!\,m!} {n!(l-1-n)!(m-1-n)!}\, 
\frac{1}{(2\pi)^{d(l+m-3-2n)}} \int \prod_{i=1}^{l-1-n} \!\!\! \rmd^d p_i \, \tphi(p_i) \,
\int \prod_{j=1}^{m-1-n} \!\!\! \rmd^d q_j \, \tphi(q_j) \,  \nn \\
\noalign{\vskip -4pt}
&{}\qquad \times \delta^d\big ( p_{(l-1-n)} + q_{(m-1-n)}\big ) \,
F_n(p_1,\dots,p_{l-1-n};q_1,\dots,q_{m-1-n} ) \, , 
\label{YOO}
\end{align}
where, with $p_{(l-1-n)} + q_{(m-1-n)}=0$,
\begin{align}
&{} F_n(p_1,\dots,p_{l-1-n};q_1,\dots,q_{m-1-n} ) \nn \\
&{} = \frac{1}{(2\pi)^{dn}} \int \prod_{h=1}^n \rmd^d r_h \, G(r_h) \, 
O\big (p_1,\dots,p_{l-1-n},r_n,\dots,r_1,- p_{(l-1-n)} - r_{(n)}\big  ) \nn \\
\noalign{\vskip -4pt}
&\quad \quad{}\times 
O'\big (q_1,\dots,q_{m-1-n},-r_n,\dots,-r_1,- q_{(m-1-n)} - r_{(n)} \big ) \,
K'\big ( (p_{(l-1-n)} + r_{(n)})^2 \big ) \, .
\end{align}
For the cases of interest here
\begin{align}
& \hskip -0.5cm  \OO = \OO_{l0}\, , \quad \OO'= \OO_{m0} \, , \nn \\
& \hskip -0.5cm   F_n(p_1,\dots,p_{l-1-n};q_1,\dots,q_{m-1-n} ) 
= \rho_n \big ( p_{(l-1-n)}{\!}^2 \big )\, , \nn \\
& \hskip -0.5cm  \OO = \OO_{l0}\, , \quad \OO'= \OO_{m2} \, , \nn \\
& \hskip -0.5cm  F_n(p_1,\dots,p_{l-1-n};q_1,\dots,q_{m-1-n} ) 
= {\textstyle {\sum_{j=1}^{m-1-n} \! q_j{\!}^2}} \, 
\rho_n \big ( p_{(l-1-n)}{\!}^2 \big ) + \tau_n \big ( p_{(l-1-n)}{\!}^2 \big )\, , 
\label{OO2}
\end{align}
with
\begin{subequations}
\begin{align}
\rho_n(p^2) = {}&  \frac{1}{(2\pi)^{dn}} \int \prod_{h=1}^n \rmd^d r_h \, G(r_h) \, 
K'\big ( (p + r_{(n)})^2 \big ) \, , \label{rta} \\
\tau_n(p^2) = {}&  \frac{1}{(2\pi)^{dn}} \int \prod_{h=1}^n \rmd^d r_h \, G(r_h) \, 
\big ( n \, r_n{\!}^2 + (p + r_{(n)})^2 \big ) \, K'\big ( (p + r_{(n)})^2 \big ) \, .
\label{rtb}
\end{align}
\end{subequations}
With the aid of \eqref{YOO} and \eqref{OO2} we may then write
\begin{align}
\frac{1}{(2\pi)^d} \int \rmd^d p \, K'(p^2) \, \frac{\delta \OO_{l0}}{\delta \tphi(p)}
\frac{\delta \OO_{m0} }{\delta \tphi(-p)} =  \sum_k \big ( 
\OO_{k0} \,  C_{klm} + \OO_{k2} \,  {\tilde C}_{klm} + \dots \big ) \, , \nn \\
 \frac{1}{(2\pi)^d}\int \rmd^d p \, K'(p^2) \, \frac{\delta \OO_{l0}}{\delta \tphi(p)}
\frac{\delta \OO_{m2} }{\delta \tphi(-p)} = \sum_k \big (
\OO_{k0} \,  D_{klm} + \OO_{k2} \,  {\tilde D}_{klm} + \dots \big ) \, , 
\end{align}
for
\begin{align}
C_{klm} = {}& \frac{l!\,m!}{(s-1-k)!(s-l)!(s-m)!} \, \rho_{s-1-k}(0) \, , \nn \\
{\tilde C}_{klm} = {}& \frac{l!\,m!}{(s-1-k)!(s-1-l)!(s-1- m)!} \, 
\rho'{\!}_{s-1-k}(0) \, , \nn \\
D_{klm} = {}& \frac{l!\,(m-2)!}{(s-1-k)!(s-l)!(s-m)!} \, \tau_{s-1-k}(0) \, , \nn \\
{\tilde D}_{klm} = {}& \frac{l!\,(m-2)!}{(s-1-k)!(s-1-l)!(s- m)!} \, 
\big ( (k-1) \rho_{s-1-k}(0) + \tau'{\!}_{s-1-k}(0) \big ) \, ,
\label{CD}
\end{align}
which are non zero so long as $s=\half(k+l+m) = k+1,k+2,\dots$.

The truncation of the full RG equation \eqref{full} corresponding to the
derivative expansion as considered in Section \ref{sec:2od} is obtained by
writing
\be
S = \textstyle{\sum_k} \big ( a_k \, \OO_{k0} + b_k \, \OO_{k2} \big ) \, ,
\ee
and then reducing \eqref{full} to
\begin{subequations}
\begin{align}
{\dot a}_k = {}& \big ( d - \half ( d - 2 +  \eta) k \big ) \, a_k +
\textstyle{\sum_{l,m} \big ( C_{klm} \, a_l a_m + 2 D_{klm} \, a_l b_m} \big ) \, , 
\label{aba} \\
{\dot b}_k = {}& \big ( d -2 - \half ( d - 2 +  \eta) k \big ) \, b_k  - \half \eta \, 
\delta_{k2} +  \textstyle{\sum_{l,m} \big ( {\tilde C}_{klm} \, a_l a_m + 
2 {\tilde D}_{klm} \, a_l b_m} \big ) \, .
\label{abb}
\end{align}
\end{subequations}
These equations are arbitrary up to the rescalings
\begin{align}
a_k \to a_k/\alpha_k \, , \ b_k \to b_k/\beta_k \, ,
\quad C_{klm} \to {}& C_{klm}\alpha_l\alpha_m/\alpha_k \, , \quad 
D_{klm} \to  D_{klm}\alpha_l\beta_m/\alpha_k \, , \nn \\
{\tilde C}_{klm} \to {}& {\tilde C}_{klm}\alpha_l\alpha_m/\beta_k \, , \quad  
{\tilde D}_{klm} \to   {\tilde D}_{klm}\alpha_l\beta_m/\beta_k \, , 
\end{align}
for any $\alpha_k,\beta_k$ so long as $\beta_2=1$ because of the inhomogeneous 
$\eta$ term in \eqref{abb}. 
Deferring further discussion to later we may impose $b_2=0$. Letting
$\alpha_k= u^{\frac{1}{2}k-1}v$, $\beta_k= u^{\frac{1}{2}k-1}$ then these rescalings 
correspond to changes in the cut off dependent functions in \eqref{CD} of
the form
\be
\rho_n(0) \to u^n v \, \rho_n(0) \, , \quad \rho'{\!}_n(0) \to u^n v^2 \, 
\rho'{\!}_n(0) \, , \quad \tau_n(0) \to u^n \, \tau_n(0) \, , \quad 
\tau'{\!}_n(0) \to u^n v \, \tau'{\!}_n(0) \, .
\label{amb}
\ee
Critical exponents calculated from \eqref{aba},\eqref{abb} must be independent of such 
transformations.

The analysis of \eqref{aba},\eqref{abb} in the $\vep$-expansion, with $d$ as in \eqref{dee},
is very similar to the previous discussion in the context of the derivative expansion.
At a fixed point a consistent solution is obtained with
\be
a_{*k} = \vep \, a_* \delta_{k\, 2n} + \vep^2 \, a_{*k}{\!}^{(2)} + {\rm O}(\vep^3) \, ,
\quad b_{*k} = \vep^2 \, b_{*k}{\!}^{(2)} + {\rm O}(\vep^3) \, , \qquad
\eta = \vep^2 \, \eta^{(2)} + {\rm O}(\vep^3)  \, ,
\ee
where
\begin{align}
a_* = {}& - \frac{n-1}{C_{2n\,2n\,2n}} \, , \qquad \quad 
\eta^{(2)} = 2a_*{\!}^2 {\tilde C}_{2\,2n\,2n} \, , \nn \\
b_{*2p}{\!}^{(2)} = {}& \frac{n-1}{2(p-1)} \, a_*{\!}^2 {\tilde C}_{2p\,2n\,2n}\, , \
p \ne 1 \, , \qquad 
a_{*2p}{\!}^{(2)} = \frac{n-1}{2(p-n)} \, a_*{\!}^2 {C}_{2p\,2n\,2n} \, , \ p\ne n \, ,
\nn \\
a_{*2n}{\!}^{(2)} = {}& - \frac{2a_*}{n-1} \Big ( {\textstyle {\sum_{p\ne n}}}
C_{2n\,2n\,2p}\, a_{*2p}{\!}^{(2)} +  
{\textstyle {\sum_{p\ne 1}}} D_{2n\,2n\,2p}\, b_{*2p}{\!}^{(2)}
\Big ) \, .
\end{align}
Hence with \eqref{CD} this gives
\be
\eta^{(2)} = 4 (n-1)^2 \frac{n!^6}{(2n)!^3}\, \frac{2n-1}{n} \,
\frac{\rho'{\!}_{2n-1}(0)} {\rho_{n-1}(0)^2} \, .
\label{eta22}
\ee
In general $\rho_n$, and also $\tau_n$, as defined in \eqref{rta} and \eqref{rtb}
depend on the cut off function but as shown in Appendix \ref{appB} there are 
certain universal quantities which are related to logarithmic divergences. In particular
\be
\frac{\rho'{\!}_{2n-1}(0)} {\rho_{n-1}(0)^2} \bigg |_{d=d_n} = \frac{n}{2n-1} =
\frac{d_n}{d_n+2} \, , 
\label{part}
\ee
so that \eqref{eta22} is in exact accord with \eqref{etaone}.

In the neighbourhood of a fixed point the solutions of \eqref{aba}, \eqref{abb}
are written as
\be
a_k(t) = a_{*k} + e^{\lambda t} f_k \, , \qquad
b_k(t) = b_{*k} + e^{\lambda t} g_{k+2(n-1)} \, , 
\ee
for $f_k, g_{k+2(n-1)}$ small.
The critical exponents are then determined by the linear eigenvalue equation
\begin{equation}
\frac{2n-k}{n-1} \, 
\lt(\begin{array}{cc}f_{k}  \\  g_k \\ \end{array}\rt) + \sum_l
\lt(\begin{array}{cc}\M_{11,kl} & \M_{12,kl}  \\ \M_{21,kl} & 
\M_{22,kl} \\ \end{array}\rt) 
\lt(\begin{array}{cc}f_{l}  \\  g_l \\ \end{array}\rt) = \lambda
\lt(\begin{array}{cc}f_{k}  \\  g_k \\ \end{array}\rt) \, , 
\label{crit}
\end{equation}
for
\begin{align}
\M_{11,kl} = {}& \big ( \vep (\half k - 1) - \half k\eta \big ) \delta_{kl}
+ 2 {\textstyle{\sum_m}} \big ( C_{klm} \, a_{*m} + D_{klm}\, b_{*m} \big ) \, , \nn \\
\M_{22,kl} = {}& \big ( \vep (\half k - n) - (\half k - n+1)\eta \big ) \delta_{kl} 
+ 2 {\textstyle{\sum_m}} {\tilde D}_{k-2(n-1)\ m \ l-2(n-1) }\, 
a_{*m}^{\vphantom g}  \, , \nn \\
\M_{12,kl} = {}& 2 {\textstyle{\sum_m}}D_{km\ l-2(n-1)}\, a_{*m}^{\vphantom g} \, , \nn \\
\M_{21,kl} = {}& 2 {\textstyle{\sum_m}} \big ( {\tilde C}_{k-2(n-1)\ lm} \, 
a_{*m}^{\vphantom g} + {\tilde D}_{k-2(n-1)\ lm}\, b_{*m}^{\vphantom g} \big ) \, . 
\end{align}
To first order in $\vep$ we may let $a_{*m} \to \vep \, a_* \delta_{m\, 2n}$ and
$b_{*m} \to 0$. Since
\be
D_{k \, 2n \, k-2(n-1)} = \frac{2n}{k+1-2n} \, \tau_0(0) = 0 \, ,
\ee
as from \eqref{rtb} $\tau_0(p^2) = p^2 K'(p^2)$, then $\M_{12,kl}$ is ${\rm O}(\vep^2)$, 
whereas otherwise  $\M_{11,kl},\M_{21,kl}$ and $\M_{22,kl}$ are ${\rm O}(\vep)$. 
Hence the first order critical exponents from \eqref{crit} are
\begin{align}
\lambda_{1,k}{\!}^{(1)} = {}& \half k - 1 + 2 a_* \, C_{kk\, 2n}
= \half k - 1 - 2(n-1)\frac{n! \,k!}{(2n)! \, (k-n)!} \, , \nn \\
\lambda_{2,k}{\!}^{(1)} = {}& \half k - n + 2 a_* \, 
{\tilde D}_{k-2(n-1)\ 2n \ k-(2(n-1)} \nn \\
= {}& \half k - n  -  2(n-1)\frac{n! \,(k-2n)!}{(2n)! \, (k-3n+1)!} \bigg (
k-2n+1 + n \, \frac{\tau'{\!}_{n-1}(0)}{\rho_{n-1}(0)} \bigg ) \, .
\label{lam12}
\end{align}
Just as in \eqref{part}
\be
\tau'{\!}_{n-1}(0) \big |_{d=d_n} = 0 \, ,
\label{tzero}
\ee
so that \eqref{lam12} is in exact agreement with the perturbative results
\eqref{expk} and \eqref{expk2}.

\section{Modified Derivative Expansion}\label{mod}

The results of Section \ref{Scaling} can be rewritten in a form which is close
to the derivative expansion. To achieve this it is necessary to make specific
choices of the cut off dependent quantities which appear in \eqref{CD} and which
are arbitrary up to the freedom exhibited in \eqref{amb}. It is crucial of course that
the cut-off function independent results in \eqref{part} and \eqref{tzero}, as well as
$\tau_0(0)=0$, should be satisfied. To this end we choose
\be
\rho_n(0) = -1 \, , \qquad \rho'{\!}_n(0) = \frac{d}{d+2} \, , \qquad
\tau_n(0) = \A n \,  , \qquad \tau'{\!}_n(0) = 0 \, ,
\ee
where $\A$ is arbitrary. 
With these choices, and with \eqref{eq:Nm} and \eqref{eq:Gklm}, then \eqref{CD}
gives
\begin{align}
C_{klm} = {}& -2^{\frac{1}{2}(k-l-m)+1} \, \frac{1}{N_k} \, lm\, G_{k\, l-1\, m-1}
\, , \nn \\
{\tilde C}_{k+2\, lm} = {}& \frac{d}{d+2} \, 2^{\frac{1}{2}(k-l-m)+2} \, 
\frac{1}{N_k} \, l(l-1)m(m-1) \, G_{k\, l-2\, m-2} \, , \nn \\
D_{kl\, m+2} = {}& \A\, 2^{\frac{1}{2}(k-l-m)+1} \, \frac{1}{N_k} \, 
l(l-1) G_{k\, l-2\, m}   \, , \nn \\
{\tilde D}_{k+2\,l\,m+2} = {}& -2^{\frac{1}{2}(k-l-m)+1} \, \frac{1}{N_k} \, 
\big ( lm\, G_{k\, l-1\, m-1} + l(l-1)\,  G_{k\, l-2\, m} \big )  \, .
\label{CDG}
\end{align}
If we now define
\be
{\tv}(x) = \sum_k a_k \, 2^{-\frac{1}{2}k} H_k(x) \, , \qquad
{\tz}(x) = \sum_k b_{k+2} \, 2^{-\frac{1}{2}k} H_k(x) \, ,
\ee
then the truncated RG equations \eqref{aba} and \eqref{abb} are equivalent, subject to
requiring non singular solutions for all $x$, to the coupled differential equations
\begin{subequations}
\begin{align}
{\dot \tv} = {}& \big ( d + \half(d-2+\eta)D \big ) \tv - \half \, \tv'{}^2 + 
\A \, \tv'' \tz \, , \\
{\dot \tz} = {}& \big ( -\eta  + \half(d-2+\eta)D \big ) \tz
- \tv' \tz' - \tv''\tz - \half \eta + \frac{d}{4(d+2)} \, \tv''{}^2 \, .
\end{align}
\end{subequations}
With the redefinitions
\be
v = \frac{2}{d-2+\eta} \, \tv \, , \qquad z = 2 \tz \, ,
\ee
and, with $K,L$ defined as in  \eqref{KLAB}, then these become
\begin{subequations}
\begin{align}
\frac{2}{d-2+\eta}\, {\dot v} = {}& ( D+K ) v - \half \, v'{}^2 +
{\tilde \A} \, v'' z \, , \label{reva} \\
\frac{2}{d-2+\eta} \, {\dot z} = {}&  ( D - L  ) z
- v' z' - v''z - L  + \half {\tilde \B} \, v''{}^2 \, ,
\label{revb}
\end{align}
\end{subequations}
for
\be
{\tilde \A} = \frac{2}{d-2+\eta} \, \A \, , \qquad 
{\tilde \B} = \half (d-2+\eta) \, \frac{d}{d+2} \, .
\ee

The results \eqref{reva} and \eqref{revb} are very similar to \eqref{eq:2orga}
and \eqref{eq:2orgb}, although there is no linear $z$ term in \eqref{reva}
and the coefficient ${\tilde \B}$  is determined in \eqref{revb}. As a consequence
the coefficient $\ck_n$ in the leading order solution \eqref{vnn} is unchanged
from \eqref{eq:kn}. Furthermore following the same discussion as in Sections
\ref{sec:2od} and \ref{sec:exponentssecondorder} gives the correct values
for $\eta$ to \Oee. Thus instead of \eqref{del0}
\begin{equation}
\Delta^{(0)}=\lt(\begin{array}{cc}-D -2n & 0  \\
0 & -D  \\ \end{array}\rt) \, ,
\end{equation}
and replacing \eqref{del1}
\be
\Delta^{(1)}=\lt(\begin{array}{cc}-(n-1)^2+\ck_nH_{2n}{\!}'\frac{\rmd}{\rmd x} &
- {\tilde \A}\,  \ck_n H_{2n}{\!\!}'' \\
- \tilde{\B}\, \ck_nH_{2n}{\!\!}'' \frac{\rmd^2}{\rmd x^2} &
\ck_nH_{2n}{\!\!}'\frac{\rmd}{\rmd x}+ \ck_nH_{2n}{\!\!}'' \\
\end{array}\rt)\, .
\end{equation}
It is then easy to see that this ensures the correct \Oe\ result instead of
\eqref{eq:splitting2} as well as preserving \eqref{eq:splitting1}.

In fact it is easy to verify that \eqref{exact1}, \eqref{exact2} and \eqref{exact3}
still give exact eigenfunctions and eigenvalues for the linearised 
perturbations \eqref{reva} and \eqref{revb} about fixed points $v_*,z_*$.
These considerations may ensure that \eqref{reva} and \eqref{revb} have a greater
chance of predictive success when they are analysed without using the $\vep$-expansion.
Of course setting $z,\eta$ to zero in  \eqref{reva} reduces it to just the LPA.

\section{Conclusion}

The status of the derivative expansion for exact RG flow equations is not entirely
clear. In some respects it may be similar to effective field theories describing
the large distance or low energy aspects of more fundamental theories. Having identified
the relevant degrees of freedom and appropriate symmetries an effective lagrangian is
constructed in terms of all symmetric scalars formed from the basic fields up to some
scale dimension so as to reproduce physical amplitudes as far as contributions of the
form $(E/\Lambda)^p$ for some $p$ where $E$ is a physical  energy scale and $\Lambda$ a 
cut off \cite{Effect}. The couplings which appear in the effective lagrangian can
in principle be determined by matching the predictions of the effective theory
with the fundamental theory for some specific physical amplitude. In a somewhat similar
fashion a derivative expansion generates terms in the differential flow equations 
whose coefficients appear to depend on the cut off function and so are essentially
arbitrary. A possible resolution is to match the results to those coming from
the $\vep$-expansion for $\vep\to 0$, although the approximate flow equations
may then be used for general $d$.
The results in this paper essentially show how this can be achieved to \Oe\ 
and to include thereby the universal aspects of all two vertex Feynman
graphs. It would be interesting although non trivial to extend this to three
vertex graphs. The results in Appendix \ref{PVVV} show how at this order various
transcendental numbers arise which make achieving this for different multicritical
points simultaneously hard to achieve.

An important issue in using exact RG equations is to determine which solutions 
are physically relevant and give results independent of the particular RG
equation or the detailed cut off function as an infra-red fixed point is approached
and we may take $\Lambda\to \infty$. 
This question becomes more significant in approximation schemes when the symmetries 
of the original exact RG  equation are no longer maintained and spurious solutions 
and critical exponents may be generated. The Polchinski RG equations for a fixed
point action $S_*[\vphi]$ has an exactly marginal operator with zero critical exponent. 
This ensures that there is in general a line of physically equivalent fixed points 
$S_*[\vphi,a]$ depending on a parameter $a$. The exact marginal operator, which is 
constructed in detail in Appendix \ref{appD}, corresponds to an infinitesimal change in 
the scale of the field $\vphi$ under which the functional integral is invariant 
(conventionally the kinetic term in the action may be normalised to one
but this is not essential, the physical couplings need only be redefined appropriately), 
for a further discussion see \cite{Com}. In the perturbative context the presence of 
such a marginal operator was demonstrated after \eqref{anom} and is a property 
of the \Oe\ results in \eqref{expk2} for $k=2n$. The presence of the irrelevant gauge
parameter $a$ is in general necessary for the RG equations to determine $\eta$.

If the symmetry under rescaling of the fields were to be
maintained in a derivative expansion it would imply that critical
exponents should be independent of $z_*(0)$ \cite{MorrisD}. In the derivative
expansion results obtained here in \eqref{eq:2ov}, \eqref{eq:2oz}, or the corresponding 
equations from \eqref{reva}, \eqref{revb}, there is a relation in general
between $z_*(0)$ and $\eta$ so that $\eta$ is not determined unless $z_*(0)$ is
fixed. At lowest order in $\vep$ as in \eqref{eq:2oz2} the dependence on $z_*(0)$
disappears. Imposing $z_*(0)=0$ makes the equations well defined but is not a 
necessary requirement in general. 
The marginal operator constructed in Appendix \ref{appD} involves an integration over 
$\tphi(q)$ for all $q$ and so approximations such as the derivative approximation 
emphasising low $q$ fail to maintain the exact zero value for the critical exponent. 
Nevertheless preserving as far as possible the presence of a marginal operator is then
a potential further constraint on solutions of exact RG equations in the
derivative approximation \cite{Rie} which may be used to restrict cut off dependence.

Finally we note that the LPA has desirable features which are absent in any
straightforward fashion in the derivative expansion. As shown in \eqref{newop}
the operator determining critical exponents for the LPA can be recast in self
adjoint form. Related to this is the fact that for the LPA it is possible to
construct a $C$-function from which the RG flow equations can be obtained \cite{CFun}
and that the equations can be written as a gradient flow \cite{Haagensen,Zum}.
Whether this is true more generally remains to be demonstrated.

\vfil\eject
\appendix\section{Further Perturbative Calculations}\label{PVVV}

The results in Section \ref{sec:pertex} may be extended to the the next order
${\rm O}(V^3)$ in a similar fashion and we obtain some results here. The 
one particle irreducible contributions to $W$ are, with the same notation as \eqref{W1}, given by
\begin{align}
W_2 = {}& -\frac{1}{6} \sum_{r,s,t \ge 1} \frac{1}{r!s!t!} 
\int \rmd^d x_1 \, \rmd^d x_2 \, \rmd^d x_3 \, 
V^{(r+t)}(\vphi_1)V^{(r+s)}(\vphi_2)V^{(s+t)}(\vphi_3) \nn\\
\noalign{\vskip -12pt}
& \hskip 6cm {}\times  G_0(x_{12})^r G_0(x_{23})^s G_0(x_{31})^t \nn \\
&{} - \frac{1}{2} \sum_{r,s \ge 2} \frac{1}{r!s!}
\int \rmd^d x_1 \, \rmd^d x_2 \, \rmd^d x_3 \,
V^{(r)}(\vphi_1)V^{(r+s)}(\vphi_2)V^{(s)}(\vphi_3) \, G_0(x_{12})^r G_0(x_{23})^s \nn \\
&{} + \sum_{r \ge 2} \frac{1}{r!} \int \rmd^d x_1 \, \rmd^d x_2 \,
V^{(r)}(\vphi_1) \, G_0(x_{12})^r \, V^{(r)}_{{\rm{c.t.}1}}(\vphi_2) \, , 
\label{W2}
\end{align}
where $V_{{\rm{c.t.}1}}$ is determined by the first term in \eqref{Lct}.
This removes subdivergencies arising in \eqref{W2} for $r,s,t=n$. If we restrict
$r,s,t<2n-1$  no further subtractions are necessary.

The divergencies coming from the second term in \eqref{W2} are easily obtained
since if $\R$ is the usual operation defining a finite part, so that from
\eqref{Gdiva}
\be
\R \big (G_0(x)^n \big ) = G_0(x)^n -  \frac{2}{\vep} \, \frac{1}{(4\pi)^n} \,
\Gamma \Big( \frac{1}{n-1}\Big)^{n-1} \, \d^d(x) \, ,
\ee
then in this term the finite part is given by
\be
\R \big ( G_0(x_{12})^n G_0(x_{23})^n \big ) 
= \R \big ( G_0(x_{12})^n \big ) \, \R \big ( G_0(x_{23})^n \big ) \, ,
\ee
so that the divergent pole terms are given by
$G_0(x_{12})^n G_0(x_{23})^n - \R \big ( G_0(x_{12})^n G_0(x_{23})^n \big )$.

For the first term in \eqref{W2} there is an overall divergence for $d=d_n$
when $r+s+t=2n$. To analyse this we make use of the Mellin-Barnes representation
\cite{Davy}
\begin{align}
& \int \rmd^d x_1 \, \rmd^d x_2 \, e^{i k_1\cdot x_1 + k_2\cdot x_2} \, 
G_0(x_{12})^r G_0(x_{2})^s G_0(x_{1})^t \nn \\
&{} = \frac{1}{(4\pi)^{r+s+t}} \,
\frac{\Gamma(\nu)^{r+s+t}}{\Gamma(r\nu)\Gamma(s\nu)\Gamma(t\nu)} \,
\frac{1}{\Gamma((r+s+t-1)\nu - 1)}  \, 
J_{rst}\bigg ( \frac{k_1^2}{k_3^2},\frac{k_2^2}{k_3^2} \bigg ) \,  
\bigg ( \frac{k_3^2}{4\pi} \bigg )^{(r+s+t-2)\nu - 2} ,\nn \\
& J_{rst} (u,v) =  \frac{1}{(2\pi i)^2}
\int_{\gamma-i \infty}^{\gamma+ i \infty} \!\!\!\!\!\!\!\! \rmd y
\int_{\gamma'-i \infty}^{\gamma'+ i \infty} \!\!\!\!\!\!\!\! \rmd z \
\Gamma(-y)\, \Gamma(-z) \nn \\
\noalign{\vskip -1pt}
& \hskip 3cm {} \times
\Gamma\big ((r+s-1)\nu -1 -y\big ) \, \Gamma\big ((r+t-1)\nu-1-z\big ) \nn \\
\noalign{\vskip -1pt}
& \hskip 3cm {} \times
\Gamma\big (y+z+1 - (r-1)\nu\big ) \, 
\Gamma\big (y+z+2 -(r+s+t-2)\nu \big ) \, u^y v^z \, ,
\label{GGGJ}
\end{align}
where $k_3 = - k_1 - k_2$ and $\gamma,\gamma'$ are chosen that the poles
in $y,z$ are on the opposite side of the contours from those in $y+z$. The 
functions $J_{rst} (u,v)$ satisfy various symmetry relations, in particular
\be
J_{rst}(u,v) = v^{(r+s+t-2)\nu-2} J_{srt} (u/v,1/v) = 
u^{(r+s+t-2)\nu-2} J_{tsr} (1/u,v/u) = J_{rts} (v,u) \, ,
\ee
which are necessary to ensure that  \eqref{GGGJ} is symmetric under permutations
of $k_1,k_2,k_3$ and also $s,t,r$. For $d=d_n- \vep$ as in \eqref{dee}, the 
poles in $\vep$, reflecting divergences
of relevance here, arise only from the residues of the poles at $y,z=0$.
For $r,s,t\ne n, \ r+s+t=2n$ there is then a simple $\vep$-pole arising from
$\Gamma\big (2 -2(n-1)\nu \big )$ since $\nu = \frac{1}{n-1}-\half \vep$ 
which gives 
\begin{align}
G_0(x_{12})^r G_0(x_{23})^s G_0(x_{31})^t &\, \big |_{r+s+t=2n,r,s,t\ne n} 
\sim \frac{1}{\vep} \, \frac{1}{(4\pi)^{2n}} \,
\Gamma \Big( \frac{1}{n-1}\Big)^{2n-1} K_{rst} \,
\d^d(x_{12}) \d^d(x_{13})\, , \nn \\
& K_{rst} = \frac{\Gamma \big( \frac{n-r}{n-1}\big)\, \Gamma 
\big( \frac{n-s}{n-1}\big)\, \Gamma \big( \frac{n-t}{n-1}\big)}
{\Gamma \big( \frac{r}{n-1}\big)\, \Gamma \big( \frac{s}{n-1}\big)\,
\Gamma \big( \frac{t}{n-1}\big)} \, . 
\label{G2div}
\end{align}
For $r=n, \ s+t =n$ the $y,z=0$ residues in \eqref{GGGJ} have a double
pole in $\vep$ from $\Gamma\big (1-(n-1)\nu \big )$ as well as
$\Gamma\big (2 -(2n-2)\nu \big )$. Expanding in $\vep$ gives
\begin{align}
\R\big ( G_0(x_{12})^n\big )\,& G_0(x_{23})^s G_0(x_{31})^t \, 
\big |_{s+t=n} \nn \\
\sim {}& \frac{1}{\vep^2} \, \frac{1}{(4\pi)^{2n}} \,
\Gamma \Big( \frac{1}{n-1}\Big)^{2n-2} \big ( - 2 + (n-1)^2 \vep
+ (n-1)L_{st} \, \vep \big ) \, \d^d(x_{12}) \d^d(x_{13})\, , \nn \\
& L_{st} = \psi \Big( \frac{1}{n-1}\Big) - \psi \Big( \frac{s}{n-1}\Big)
-\psi \Big( \frac{t}{n-1}\Big) + \psi(1) \, , \ s+t=n \, .
\label{G2divb}
\end{align}
With the aid of \eqref{G2div} and \eqref{G2divb} then the pole terms in
\eqref{W2} require
\begin{align}
V_{\rm{c.t.}2}(\phi)  = {}& - \frac{1}{\vep} \,  \frac{1}{(4\pi)^{2n}} \,
\Gamma \Big( \frac{1}{n-1}\Big)^{2n-1} \, \frac{1}{6} 
\sum_{\genfrac{}{}{0pt}{}{r,s,t\ge 1,r,s,t \ne n}{r+s+t=2n}} \frac{K_{rst}}{r!s!t!} \
V^{(r+t)}(\phi)V^{(r+s)}(\phi)V^{(s+t)}(\phi) \nn \\
{}& + \frac{1}{\vep^2} \,  \frac{1}{(4\pi)^{2n}} \,
\Gamma \Big( \frac{1}{n-1}\Big)^{2n-2} \, \frac{1}{n!}
\sum_{\genfrac{}{}{0pt}{}{s,t\ge 1}{s+t=n}} \frac{1}{s!t!}\, \big ( 1 - \half (n-1)^2 \vep
- \half (n-1) L_{st} \, \vep \big ) \nn \\
\noalign{\vskip -12pt}
&\hskip 6.5cm{}\times V^{(n)}(\phi)V^{(n+s)}(\phi)V^{(n+t)}(\phi) \nn \\
{}& + \frac{2}{\vep^2} \,  \frac{1}{(4\pi)^{2n}} \,
\Gamma \Big( \frac{1}{n-1}\Big)^{2n-2} \, \frac{1}{n!^2}\,
V^{(n)}(\phi)^2\, V^{(2n)}(\phi) \, .
\label{V2ct}
\end{align}
The double poles are in accord with standard RG equations from \eqref{RGdef}
\be
\bigg ( \vep - {\hat \beta}^V \! \cdot \frac{\pr}{\pr V} 
+ {\hat\gamma}_\phi \, \phi \frac{\pr}{\pr \phi} \bigg ) \big ( V(\phi)
+ V_{\rm{c.t.}}(\phi) \big ) = 0 \, ,
\ee
since we have at this order with \eqref{betaV}
\be
\beta_2^V(\phi) - \gamma_{\phi,1} \, \phi V'(\phi) +  
\beta_1^V \! \cdot \frac{\pr}{\pr V} \, V_{\rm{c.t.}1}(\phi) = 2(n-1)\vep \,
V_{\rm{c.t.}2}(\phi) \, .
\ee
{}From \eqref{V2ct} we then obtain
\begin{align}
\beta^V_{2}(\phi)  = {}& - \frac{1}{3}(n-1) \,  \frac{1}{(4\pi)^{2n}} \,
\Gamma \Big( \frac{1}{n-1}\Big)^{2n-1} \!\!\!\!
\sum_{\genfrac{}{}{0pt}{}{r,s,t\ge 1,r,s,t \ne n}{r+s+t=2n}} \frac{K_{rst}}{r!s!t!} \
V^{(r+t)}(\phi)V^{(r+s)}(\phi)V^{(s+t)}(\phi) \nn \\
{}& -(n-1)^2 \,  \frac{1}{(4\pi)^{2n}} \,
\Gamma \Big( \frac{1}{n-1}\Big)^{2n-2} \, \frac{1}{n!}
\sum_{\genfrac{}{}{0pt}{}{s,t\ge 1}{s+t=n}} \frac{1}{s!t!}\, 
\big ( n-1 +L_{st} \big ) \nn \\
\noalign{\vskip -12pt}
&\hskip 7cm{}\times V^{(n)}(\phi)V^{(n+s)}(\phi)V^{(n+t)}(\phi) \nn \\
\noalign{\vskip -6pt}
{}& + \gamma_{\phi,1} \, \phi V'(\phi) \, .
\label{beta2}
\end{align}
From \eqref{beta2} as in \eqref{bgp}
\begin{align}
{\hat \beta}_2^{V_\lambda+U} (\phi) = {}& \frac{1}{(2n)!}\, {\hat \beta}_2^\lambda(\lambda) \, \phi^{2n} +  \D_{\lambda,2} U(\phi) + {\rm O}(U^2) \, , 
\label{bgp3}
\end{align}
where $\D_{\lambda,2} U(\phi) $ determines $\gamma_{k,2}$.

For a single ${\cal L}^Z$ insertion there are contributions at ${\rm O}(\lambda^2)$ which necessitate
extra terms in $V_{\rm{c.t.}2}$. These involve
\begin{align}
W^Z_2 = {}& -  \sum_{r,s \ge 1} \frac{1}{r!s!} 
\int \rmd^d x_1 \, \rmd^d x_2 \, \rmd^d x_3 \, 
V^{(r+1)}(\vphi_1)V^{(r+s)}(\vphi_2)Z^{(s)}(\vphi_3) \nn\\
\noalign{\vskip -12pt}
& \hskip 6cm {}\times  G_0(x_{12})^r G_0(x_{23})^s  \pr^2 G_0(x_{31}) \nn \\
={}& \sum_{r,s\ge 1} \frac{1}{r!s!} 
\int \rmd^d x_1 \, \rmd^d x_2 \,
V^{(r+1)}(\vphi_1)V^{(r+s)}(\vphi_2)Z^{(s)}(\vphi_2) \, G_0(x_{12})^{r+s} \, .
\label{WZ2}
\end{align}
Using \eqref{Gdiva}
\begin{align}
\Delta \beta^V_2 (\phi) =  \frac{n-1}{n!} \,\frac{2}{(4\pi)^n} \,  
\Gamma &  \Big( \frac{1}{n-1}\Big)^{n-1} \, V^{(n)}(\phi) \nn \\
\noalign{\vskip - 4pt}
& \hskip 0.4cm  {}\times \bigg ( \frac {\rmd^n} {\rmd \phi^n} \big ( V'(\phi) Z(\phi) \big ) - V^{(n+1)} Z(\phi) - V'(\phi) Z^{(n)}(\phi) \bigg ) \, .
\end{align}
From this we may obtain
\begin{align}
\gamma^{gh}_{k,2} =  {}& \lambda^2\, \frac{n-1}{n!^2} \,\frac{2}{(4\pi)^n} \,  
\Gamma  \Big( \frac{1}{n-1}\Big)^{n-1}\,\frac{1}{(2n-1)!}\,  \frac{k!}{(k-2n+1)!} \nn \\
&\hskip 3cm {}\times \bigg ( \frac{k!}{ (k-n)!} - \frac{(k-2n+1)!}{(k-3n+1)!} -\frac{(2n)!}{2\, n! } \bigg ) \, .
\label{ghgg}
\end{align}
As a special case
\be
\gamma^{gh}_{2n,2} =  \lambda^2 \frac{(2n)!}{(n-2)!\,  n!^2} \,\frac{2}{(4\pi)^n} \,  
\Gamma  \Big( \frac{1}{n-1}\Big)^{n-1} = 2n \, \beta^\lambda_1 \, ,
\ee
with $\beta^\lambda_1$ determined in \eqref{Vone}. Also
\begin{align}
\gamma^{gh}_{k,2} \big |_{n=2} ={}&  \frac{\lambda^2}{(4\pi)^2} \, \tfrac12 k(k-1)(k-2)(k-3)\, , \nn  \\
\gamma^{gh}_{k,2}  \big |_{n=3} = {}&  \frac{\lambda^2\pi}{(4\pi)^3} \, \tfrac{1}{72} k(k-1)(k-2)^2(k-3)(k-4)(k-5)\, .
\label{resghg}
\end{align}

For general $n$ it is not straightforward to analyse \eqref{beta2} further so
we content ourselves for the simplest cases of $n=2,3$ which give
\be
\beta_2^V(\phi) \big |_{n=2} = -\half \, \frac{1}{(4\pi)^4} \, V^{(2)}(\phi)
V^{(3)}(\phi)^2 + \gamma_{\phi,1} \, \phi V'(\phi) \, ,
\quad  \gamma_{\phi,1}= {\frac{1}{12}} \, \frac{\lambda^2}{(4\pi)^4} \, , 
\label{beta22}
\ee
and
\begin{align}
\beta_2^V(\phi) \big |_{n=3} = {}& \frac{1}{(8\pi)^4} \Big (
{\textstyle\frac{1}{6}} \, V^{(2)}(\phi)V^{(5)}(\phi)^2
- {\textstyle\frac{1}{12}}\pi^2\, V^{(4)}(\phi)^3
- \tfrac43\, V^{(3)}(\phi)V^{(4)}(\phi)V^{(5)}(\phi)\Big )  \nn \\
\noalign{\vskip -8pt}
&{} + \gamma_{\phi,1} \, \phi V'(\phi) \, , \qquad\qquad
\gamma_{\phi,1} = 
{\frac{1}{90}}\,  \frac{\lambda^2}{(8\pi)^4} \, .
\end{align}
Using \eqref{beta22} we may obtain \Oee\ corrections to \eqref{gstar} and \eqref{expk}
for $n=2$
\begin{align}
\frac{3\lambda_*}{(4\pi)^2} = {}& \vep + {\textstyle\frac{2}{3}}\, \vep^2 - 2 \eta \, , \qquad
\eta = {\textstyle\frac{1}{54}} \, \vep^2 + {\rm O}(\vep^3) \, , \nn \\
{\hat \gamma}_k(\lambda_*) = {}& - \half ( k - 2) \, \vep + \half k \, \eta 
+ \textstyle{\frac{1}{6}}k(k-1)(\vep-2\eta) - \textstyle{\frac{1}{18}} 
k(k-1)(k-3)\, \vep^2 + {\rm O}(\vep^3) \, .
\label{res2}
\end{align}
This agrees with standard results for $k=1,2,3$. Furthermore for $n=3$
\begin{align}
& \frac{20}{3}\, \frac{\lambda_*}{(8\pi)^2} = 2\vep + {\textstyle\frac{27}{40}}
\big (10 + \pi^2  \big ) \, \vep^2 - 3 \eta \, , \qquad
\eta = {\textstyle\frac{1}{500}} \, \vep^2 + {\rm O}(\vep^3) \, , \nn \\
& {\hat \gamma}_k(\lambda_*) = - \half ( k - 2) \, \vep + \half k \, \eta
+ \textstyle{\frac{1}{30}}k(k-1)(k-2)\big(\vep- {\textstyle\frac{3}{2}} \eta\big ) \nn \\
&\hskip 1.4cm {} - \textstyle{\frac{1}{100}} k(k-1)(k-5) \big ( k(k-2)
-\textstyle{\frac{1}{8}}k(k-4) + \textstyle{\frac{9}{16}}\pi^2(k-2) \big ) \, 
\vep^2 + {\rm O}(\vep^3) \, .
\label{res3}
\end{align}

When $k\ge 2n$ the \Oee\ results are modified due to mixing effects. Without computing
the \Oee\ terms in ${\hat \gamma}^{hh}_k(\lambda_*)$ in the matrix \eqref{anom}, using the left and right
eigenvectors for the matrix \eqref{anom3}, we have for one eigenvalue
\be
\omega_{1,k} = {\hat \gamma}_k(\lambda_*) + \Delta \omega_{1,k} + {\rm O}(\vep^3) \, , \qquad
\Delta \omega_{1,k} = \frac{{\gamma}^{gh}_{k,2}(\lambda_*) \,{\gamma}^{hg}_{k,1}(\lambda_*)}
{{\hat \gamma}^{gg}_{k,1}(\lambda_*) - {\hat \gamma}^{hh}_{k,1}(\lambda_*)} \, .
\ee
From \eqref{expk} and \eqref{expkh}
\be
{\hat \gamma}^{gg}_{k,1}(\lambda_*) - {\hat \gamma}^{hh}_{k,1}(\lambda_*)
= 2(n-1) \frac{n!}{(2n)!} \bigg ( \frac{k!}{ (k-n)!} - \frac{(k-2n+1)!}{(k-3n+1)!} -\frac{(2n)!}{2\, n! } \bigg ) \vep\, .
\ee
Using \eqref{ghgg} with \eqref{ghg1} this gives
\begin{align}
\Delta \omega_{1,k}  = \frac{1} {(2n-1)!}\,  \frac{k!}{(k-2n +1)!} \, \eta \, , 
\qquad \ \ k\ge 2n  \, , 
\end{align}
with $\eta$ as in \eqref{res2} and \eqref{res3}.  For $k=2n$
the result is consistent with $\omega_{1,2n} = {\hat \beta}^\lambda{}'(\lambda_*)$.

By considering the residues in \eqref{GGGJ} at $y=1,z=0$ and $y=0,z=1$, and
requiring $r+s+t=3n-1$, we may also determine directly higher order contributions to 
$\gamma_\phi$ although it is then necessary to include an additional counterterm for
when $r,s,t=2n-1$ in \eqref{W2}. 
Such results are omitted as they are irrelevant in the context of this paper.

\section{Verification of Vanishing of a Sum}
\label{sec:vanishingsum}

In the discussion of critical exponents in Section \ref{sec:exponentsfirstorder} 
consistency required that the sum appearing in \eqref{sumtwo}
\begin{equation}
S= \sum_{p=0}^{n}(2n-3p)\frac{(2p)!}{p!^4(n-p)!^2} \, ,
\end{equation}
should vanish.
Although in the case where it arises here $n$ is even $S=0$ for any $n$.
To show this directly we note that
\begin{equation}
S= \frac{1}{n!^2}\sum_{p=0}^{n}(2n-3p)\binom{2p}{p}\binom{n}{p}^2 \, ,
\end{equation}
where we may write
\begin{equation}
\sum_{r=0}^{p}\binom{p}{r}^2=\binom{2p}{p} \, .
\end{equation}
Hence 
\begin{align}
S= {}& \frac{1}{n!^2}\sum_{p=0}^{n}\sum_{r=0}^{p}(2n-3p)
\binom{p}{r}^2\binom{n}{p}^2\nn\\
= {} & \sum_{p=0}^{n}\sum_{r=0}^{p}(2n-3p)
\frac{1}{r!^2(p-r)!^2(n-p)!^2}\nn\\
={} &\sum_{s=0}^{n}\sum_{r=0}^{n-s}(2n-3s-3r)\frac{1}{r!^2s!^2(n-s-r)!^2}\, ,
\end{align}
where $s=p-r$. Then, setting $t=n-s-r$,
\begin{align}
S=&\sum_{\substack{r,s,t\ge 0 \\ r+s+t=n}}(2t-r-s)\frac{1}{r!^2s!^2t!^2} \, ,
\end{align}
from which it follows using symmetry of the $r,s,t$-sums that $S=0$. This
then implies \eqref{tzero}.

\section{Integrals and Cut Off Function Dependence}\label{appB}

In the discussion in section \ref{Scaling} the dependence on the cut off
function was reduced to particular integrals such as appeared in \eqref{rta}
and \eqref{rtb}. In general the presence of an arbitrary cut off function $K(p^2)$,
constrained only by $K(0)=1$ and rapid fall off for large $p^2$, ensures
that they can take any value but in special cases the integrals are identical
with the logarithmically divergent part of standard Feynman integrals and so
they have a universal form independent of any particular $K(p^2)$. 

Reinstating the cut off in the propagator which appears in \eqref{GY} so that 
\be
G_\Lambda(p) = \frac{K(p^2/\Lambda^2)}{p^2} \, ,
\ee
then \eqref{rta} can be written as
\begin{align}
\Lambda^{m(d-2)-2} \rho_m\big (p^2/\Lambda^2 \big ) = {}& \frac{1}{\Lambda^2} 
\, \frac{1}{(2\pi)^{dm}} \! \int \prod_{h=1}^m \rmd^d r_h \, 
G_\Lambda (r_h) \, K'\big ( (p + r_{(m)})^2 /\Lambda^2 \big ) \nn \\
= {}& - \frac{1}{2(m+1)} \, \Lambda \frac{\pr}{\pr \Lambda} \, \frac{1}{(2\pi)^{dm}} 
\! \int \prod_{h=1}^m \rmd^d r_h \, G_\Lambda (r_h) \, G_\Lambda (p + r_{(m)} ) \, .
\end{align}
The logarithmic divergencies present in the product of $m+1$ propagators for $d=d_n$ 
as in \eqref{dee} when $m(d_n-2) = 2+ 2l, \ l=0,1,\dots$ then generate a cut off 
independent result for $\rho_m^{(l)}(0)$.

To obtain the detailed coefficients, following \cite{Phase}, the momentum space
convolution integrals in \eqref{rta}, \eqref{rtb} are expressed in terms of 
the $x$-space propagator
\be
{\tilde G} (x) = \frac{1}{(2\pi)^d} \int \rmd^d p \, e^{-ip\cdot x} \, G(p) \, .
\ee
and, using $(1+ \half p\cdot \pr_p ) G(p) = K'(p^2)$, 
\be
-\half \big ( d-2 + x\cdot \pr_x \big )\, {\tilde G} (x) = 
\frac{1}{(2\pi)^d} \int \rmd^d p \, e^{-ip\cdot x} K'(p^2) \, .
\ee
Then \eqref{rta} becomes
\be 
\rho_m(p^2) = - \frac{1}{2} \int \rmd^d x \,  e^{ip\cdot x} \,
{\tilde G}(x)^m \, \big ( d-2 + x\cdot \pr_x \big )\, {\tilde G} (x) \, .
\ee
This gives for the Taylor expansion coefficients at $p^2=0$
\be
\rho^{(l)}_m(0) = -  e_l\, \frac{1}{2} \int \rmd^d x \, (x^2)^l \,
{\tilde G}(x)^m \, \big ( d-2 + x\cdot \pr_x \big )\, {\tilde G} (x) \, , \quad
e_l = \frac{(-1)^l}{2^{2l}l! (\half d)_l} \, ,
\ee
or with $r^2 = x^2$
\be
\rho^{(l)}_m(0) = - e_l \, \frac{S_d}{2(m+1)} \int_0^\infty \!\!\!\! \rmd r\, 
r^{d+2l-1} \big ( (m+1)( d-2) + r \pr_r \big ) \, {\tilde G}(x)^{m+1} \, , 
\ee
for $S_d = {2\pi^{\frac{1}{2}}}/{\Gamma(\half d)}$. When $d+2l = (m+1)(d-2)$ the
integrand is a total derivative and using
\be
{\tilde G}(x) \sim \frac {1}{(d-2)S_d} \, \frac{1}{r^{d-2}} \quad\hbox{as} \quad
r\to \infty \, ,
\ee
then there is only a surface term for large $r$ giving
\be
\rho^{(l)}_{(l+1)(n-1)}(0) \big |_{d=d_n} = 
- (-1)^l e_l \, \frac{n-1}{4(n+l(n-1))} \, \big ((d_n-2) S_{d_n}\big )^{-(l+1)(n-1)}\, .
\label{Div}
\ee
This result directly implies \eqref{part}. For $l=0,1$ the coefficients 
obtained in \eqref{Div} correspond exactly to the pole terms in dimensional 
regularisation in \eqref{Gdiva}, \eqref{Gdivb}.

In a similar vein from \eqref{rtb}
\begin{align} 
\tau_m(p^2) = \frac{1}{2} \int \rmd^d x \,  e^{ip\cdot x} \,
\Big ({}& m\,  {\tilde G}(x)^{m-1}\pr^2 {\tilde G}(x) \, 
\big ( d-2 + x\cdot \pr_x \big )\, {\tilde G} (x) \nn \\
\noalign{\vskip -6pt}
&{} + {\tilde G}(x)^m  \big ( d + x\cdot \pr_x \big )\, \pr^2 {\tilde G} (x) \Big )  \, ,
\end{align}
and
\be
\tau^{(l)}_m(0) = e_l \, \frac{1}{2}S_d \int_0^\infty \!\!\!\! \rmd r\,
r^{d+2l-1} \big ( m( d-2) +d + r \pr_r \big ) \, {\tilde G}(x)^m\pr^2 {\tilde G}(x)  \, .
\ee
In this case the condition for a the integrand to be a total derivative is $m(d-2)=2l$
but there is no correspond surface term as $\pr^2 {\tilde G}(x)$ vanishes more
rapidly than $r^{-d}$ as $r\to \infty$ and therefore
\be
 \tau^{(l)}_{l(n-1)}(0) \big |_{d=d_n} = 0 \, .
\ee

\section{Perturbations of Exact RG Flow Equations}\label{appD}

We here discuss perturbations of the exact RG flow equations in \eqref{full}
which may be written, neglecting $C$, more succinctly in the form
\be
\frac{\pr}{\pr t} S = ( \D_1 + \D_2 )S + S * S  - 
\eta \, \vphi \cdot K^{-1} \vphi\, ,
\label{Sflow}  
\ee
where 
\begin{align}
S * S = {}& \frac{1}{(2\pi)^d}\int \rmd^d p \, K'(p^2) \,
\frac{\delta S}{\delta \tphi(p)}\, \frac{\delta S}{\delta \tphi(-p)} \, ,\nn \\
\vphi \cdot K^{-1} \vphi = {}& \half \, \frac{1}{(2\pi)^d}\int \rmd^d p \,
K(p^2)^{-1}p^2 \tphi(p)\tphi(-p) \, , 
\label{Srg}
\end{align}
and
using the definitions \eqref{D12} save that now
\be
\D_1 = \frac{1}{(2\pi)^d}\int \rmd^d p \,
\big ( \half d + 1 -\half \eta  + p \cdot \pr_p \big )
\tphi(p) \, \frac{\delta}{\delta \tphi(p)} \, .
\ee
For a small variation $\delta  S$
\be
\frac{\pr}{\pr t} \delta S = ( \D_1 + \D_2 + \D_S ) \delta S \, , \qquad
\D_S = \frac{2}{(2\pi)^d}\int \rmd^d p \, K'(p^2) \,
\frac{\delta S}{\delta \tphi(- p)}\, \frac{\delta}{\delta \tphi(p)} \, .
\label{dSrg}
\ee
At a fixed point $S\to S_*$ with ${\dot S}_*=0$. The critical exponents 
are then defined by
\be
\big ( \D_1 + \D_2 + \D_{S_*} \big ) \OO = \lambda \OO \, ,
\label{eigen}
\ee
for $\OO$ the corresponding eigen-operator.

{}From \eqref{Sflow} and the definitions for $\D_1,\D_2,\D_{S_*}$  we easily obtain
\begin{align}
\big ( \D_1 + \D_2 + \D_{S_*} \big ) \tphi(q) = {}& \big ( \half d + 1 -\half \eta  + q \cdot \pr_q \big )
\tphi(q) +2K'(q^2) \, \frac{\delta S_*}{\delta \tphi(-q)} \, , \nn \\
\big ( \D_1 + \D_2 + \D_{S_*} \big )  \frac{\delta S_*}{\delta \tphi(-q)} 
= {}& \big ( \half d - 1 + \half \eta  + q \cdot \pr_q \big ) \frac{\delta S_*}{\delta \tphi(-q)} 
+ \eta \, K(q^2)^{-1}q^2 \tphi(q) \, .
\end{align}
Hence there are two exact solutions of \eqref{eigen}
\begin{subequations}
\begin{align}
\OO = {}& \tphi(0) +\frac{2K'(0)}{2-\eta}\, \frac{\delta S_*}{\delta \tphi(0)} \, ,  
\qquad \lambda =  \half d + 1 -\half \eta \, ,  \label{ex1} \\
\OO = {}& \frac{\delta S_*}{\delta \tphi(0)}  \, , \hskip 3.3cm \lambda =  
\half d - 1 + \half \eta \, .
\label{ex2}
\end{align}
\end{subequations}
These are identical with the results obtained in \eqref{exact2} and \eqref{exact3} using the derivative
expansion.

More generally we consider solutions of \eqref{eigen} which may be expressed as
\be
\OO_\Psi = \OO_{\Psi,1} + \OO_{\Psi,2} \, , 
\label{OPsi}
\ee
where
\be 
\OO_{\Psi,1} = \frac{1}{(2\pi)^d}\int \rmd^d q \, \bigg ( \Psi(q) 
\frac{\delta S_*}{\delta \tphi(q)} - \frac{\delta \Psi(q) }{\delta \tphi(q)}  \bigg ) \, ,\quad 
\OO_{\Psi,2} = \frac{1}{(2\pi)^d}\int \rmd^d q \, K(q^2)^{-1}q^2\,  \Psi(q) \tphi(-q) \, .
\label{O12}
\ee
For operators of this form a perturbation $\epsilon \OO_\Psi$ may be removed by a redefinition of 
$\vphi$ in the basic functional integral 
$Z=\int \rmd [\vphi] \, e^{- \vphi \cdot K^{-1} \vphi - S_*[\vphi]}$ so that $Z$ is invariant.
Such operators are termed redundant \cite{Phase}. The operator in \eqref{ex2} is of this form 
by taking $\Psi(q) \to (2\pi)^d\delta^d(q)$.

For $\OO_{\Psi,1}$ using 
\be
\D_S \frac{\delta S}{\delta \tphi(q)}  = \frac{\delta}{\delta \tphi(q)}(S * S) \, ,
\quad \Big [ \D_1 , \frac{\delta}{\delta \tphi(q)} \Big ] =
\big (q \cdot \pr_q + \half d -1 +\half \eta \big ) \frac{\delta}{\delta \tphi(q)} \, .
\ee 
we have
\begin{align}
( \D_1 + \D_2 + \D_{S_*} ) \OO_{\Psi,1} 
= {}& \OO_{\Psi_1,1} + \frac{1}{(2\pi)^d}\int \rmd^d q \, 
\Psi(q)\,  \frac{\delta}{\delta \tphi(q)} \big ( (\D_1 + \D_2)S_* + S_* * S_* \big )\nn \\
\Psi_1(q) = {}& \big ( \D_1 + \D_2 + \D_S - q \cdot \pr_q -\half d - 1 + \half \eta \big )
\Psi(q)  \, .
\end{align}
For $\OO_{\Psi,2}$
\begin{align}
& ( \D_1 + \D_2 + \D_{S_*} ) \OO_{\Psi,2}  = \OO_{\Psi_2,2} + \OO_{\Psi_3,1} \, , \nn \\
& \Psi_2(q) = \Psi_1(q) - \eta \Psi(q) + \Psi_3(q) \, , \qquad 
\Psi_3(q) =  2K(q^2)^{-1}K'(q^2)q^2\Psi(q)\, .
\end{align}
Hence using the equation for $S_*$
\begin{align}
( \D_1 + \D_2 + \D_{S_*} ) \OO_{\Psi}  = \OO_{\Psi'} \, , \qquad
\Psi'(q) = \Psi_1(q) + \Psi_3(q) \, .
\label{psiO}
\end{align}

The result \eqref{psiO} demonstrates that the operators $\{\OO_\Psi\}$ form
a closed subspace under RG flow near a fixed point. If $\OO(q)$ is a local
operator satisfying the generalisation of \eqref{eigen}
\be
\big ( \D_1 + \D_2 + \D_{S_*} \big ) \OO(q) = 
\big (q \cdot \pr_q + \lambda_\OO \big ) \OO(q) \, ,
\label{eigenq}
\ee
then taking
\be
\Psi(q) = (q^2)^m K(q^2) \OO(q) \, ,
\label{PO}
\ee
gives an eigen-operator $\OO_\Psi$ with
\be
\lambda = \lambda_\OO - \half d - 1 + \half \eta -2m \, .
\label{lnew}
\ee
If $m$ were arbitrary the eigenvalue could take any value but for locality
we require $m$ to be an integer.

The operator in \eqref{ex1} may be extended to all $q$ by considering
\be
\OO(q) = a(q^2)\, \tphi(q) + b(q^2) \, \frac{\delta S_*}{\delta \tphi(-q)}\, ,
\label{Oph}
\ee
where $a(0)=1, \, b(0)= K'(0)/(1-\half \eta)$. Imposing \eqref{eigenq} 
with $\lambda_\OO = \half d + 1 -\half \eta$ gives $a'(x) = \half \eta K(x)^{-1}b(x)$,
$K'(x) a(x) - x b'(x) = (1-\half \eta) b(x)$ which have the solutions, assuming
$\eta<2$, 
\be
a(x) = \frac{1+xb(x)}{K(x)}  \, , \qquad 
b(x) = x^{\frac{1}{2}\eta -1} K(x) \int_0^x \! u^{-\frac{1}{2}\eta} \, 
\frac{K'(u)}{K(u)^2} \ \rmd u  \, .
\ee
When $\eta=0$, $a(x)=1, \ b(x)=(K(x)-1)/x$.

With these results and using \eqref{Oph} and \eqref{PO}, with $m=0$, in \eqref{O12} and
\eqref{OPsi} gives an exactly marginal eigen-operator with $\lambda=0$.
Integrating these marginal deformations generates solutions $S_*[\vphi,a]$
for some parameter $a$ representing a line of equivalent fixed points. The various
formulae may be verified with the Gaussian solution
\be
S_*[\vphi,a] = - \half \, \frac{1}{(2\pi)^d}\int \rmd^d p \,
\frac{p^2}{K(p^2)+a} \, \tphi(p)\tphi(-p) \, .
\ee

More generally assuming the eigen-operators corresponding to $\lambda_{k,1}$
$\OO_{k,1}$ may be extended to $\OO_{k,1}(q)$ satisfying \eqref{eigenq}
then this construction determines $\OO_{k+2n-1,2}$ so that from \eqref{lnew}
\be
\lambda_{k,2} = \lambda_{k-2n+1,1} - \half d - 1 + \half \eta \, .
\ee
This is compatible with the \Oe\ perturbative results and also the modified
derivative expansion calculations described here.

\bigskip
\noindent
{\bf Acknowledgements}

\noindent
J. O'D. is grateful to C. Bervillier and T. Morris for useful correspondence.

\vfil\eject

\bibliographystyle{utphys}
\bibliography{Eps-biblio}

 \end{document}